\documentclass[12pt]{article}
\usepackage{graphicx}

\def\tamu{Mitchell Institute for Fundamental Physics and Astronomy, Department of Physics and Astronomy, Texas A\&M University College Station, TX 77845, USA}
\def\support{\footnote{Work supported by DOE grant DE-FG02-13ER42020.}}

\def\Title#1{\begin{center} {\Large #1 } \end{center}}
\def\Author#1{\begin{center}{ \sc #1} \end{center}}
\def\Address#1{\begin{center}{ \it #1} \end{center}}

\newenvironment{Abstract}{\begin{quotation}  }{\end{quotation}}
\newenvironment{Presented}{\begin{quotation} \begin{center} 
             PRESENTED AT\end{center}\bigskip 
      \begin{center}\begin{large}}{\end{large}\end{center} \end{quotation}}
\def\Acknowledgements{\bigskip  \bigskip \begin{center} \begin{large}
             \bf ACKNOWLEDGEMENTS \end{large}\end{center}}

\def\beq{\begin{equation}}
\def\eeq#1{\label{#1}\end{equation}}
\def\eeqn{\end{equation}}

\def\beqa{\begin{eqnarray}}
\def\eeqa#1{\label{#1}\end{eqnarray}}
\def\eeqan{\end{eqnarray}}

\let\bar=\overbar

\def\Dslash{\not{\hbox{\kern-4pt $D$}}}
\def\dslash{\not{\hbox{\kern-2pt $\del$}}}

\def\msb{{\bar{\ssstyle M \kern -1pt S}}}

\def\neu#1{\widetilde\chi^0_{#1}}

\newcommand{\be}{\begin{equation}}
\newcommand{\bea}{\begin{eqnarray}}
\newcommand{\eea}{\end{eqnarray}}
\newcommand{\neuo}[1]{\ensuremath{\tilde{\chi}_{#1}^0}}
\newcommand{\chip}[1]{\ensuremath{\tilde{\chi}_{#1}^+}}
\newcommand{\chim}[1]{\ensuremath{\tilde{\chi}_{#1}^-}}

\newcommand{\chpm}[1]{\ensuremath{\tilde{\chi}_{#1}^{\pm}}}
\newcommand{\chmp}[1]{\ensuremath{\tilde{\chi}_{#1}^{\mp}}}

\newcommand{\gev}  {\mbox{${\rm GeV}$}}

\newcommand{\tev}  {\mbox{${\rm TeV}$}}

\newcommand{\invfb}{\mbox{${\rm fb}^{-1}$}}

\newcommand{\lum}  {\mbox{${\cal L}$}}

\newcommand{\pt}  {\mbox{$p_{\rm T}$}}
\newcommand{\HT} {{H_{\rm T}}}

\newcommand{\met} {\mbox{${E\!\!\!\!/_{\rm T}}$}}

\newcommand{\mttend}{\mbox{$M_{\tau\tau}^{\rm end}$}}

\newcommand{\mjttpeak}{\mbox{$M_{j\tau\tau}^{\rm peak}$}}

\newcommand{\mjtend}{\mbox{$M_{j\tau}^{\rm end}$}}

\newcommand{\meffpeak}{\mbox{$M_{{\rm eff}}^{\rm peak}$}}

\newcommand{\meffbnoWpeak}{\mbox{$M_{{\rm eff}}^{(b,~\mathrm{no}\ W)\; \rm peak}$}}

\newcommand{\mjWend}{\mbox{$M_{jW}^{\rm end}$}}

\newcommand{\tbar} {\mbox{$\overline{t}$}}

\newcommand{\ttbar}{\mbox{$t\overline{t}$}}

\newcommand{\azero}{\mbox{$A_{0}$}}
\newcommand{\tanb} {\mbox{$\tan\beta$}}
\newcommand{\sinb} {\mbox{$\sin\beta$}}
\newcommand{\mzero}{\mbox{$m_{0}$}}
\newcommand{\mhalf}{\mbox{$m_{1/2}$}}

\newcommand{ \gluino}   {\mbox{$\tilde{g}$}}

\newcommand{ \usquarkL}  {\mbox{$\tilde{u}_{L}$}}
\newcommand{ \usquarkR}  {\mbox{$\tilde{u}_{R}$}}

\newcommand{ \sbottomone}{\mbox{$\tilde{b}_{1}$}}

\newcommand{ \sbottomtwo}{\mbox{$\tilde{b}_{2}$}}

\newcommand{ \stopone}  {\mbox{$\tilde{t}_{1}$}}

\newcommand{ \stoptwo}  {\mbox{$\tilde{t}_{2}$}}

\newcommand{ \seleR}    {\mbox{$\tilde{e}_{R}$}}

\newcommand{ \seleL}    {\mbox{$\tilde{e}_{L}$}}

\newcommand{ \stauone}  {\mbox{$\tilde{\tau}_{1}$}}

\newcommand{ \stautwo}  {\mbox{$\tilde{\tau}_{2}$}}

\newcommand{ \schionezero }{\mbox{$\tilde{\chi}_{1}^{0}$}}
\newcommand{ \schitwozero }{\mbox{$\tilde{\chi}_{2}^{0}$}}
\newcommand{ \schithreezero }{\mbox{$\tilde{\chi}_{3}^{0}$}}
\newcommand{ \schifourzero }{\mbox{$\tilde{\chi}_{4}^{0}$}}
\newcommand{ \schionepm }{\mbox{$\tilde{\chi}_{1}^{\pm}$}}

\newcommand{ \schitwopm }{\mbox{$\tilde{\chi}_{2}^{\pm}$}}

\newcommand{ \alpgen } {{\tt ALPGEN}}
\newcommand{ \pythia } {{\tt PYTHIA}}
\newcommand{ \isasugra } {{\tt ISASUGRA}}

\newcommand{ \pgs }    {{\tt PGS4}}
\newcommand{ \darksusy }    {{\tt darkSUSY}}
\newcommand{ \madgraph } {{\tt MADGRAPH5}}

\newcommand{ \DMrelic }{\mbox{$\Omega_{\schionezero}h^{2}$}}

\begin{document}
\begin{titlepage}

\vfill
\Title{Dark Matter Searches at Accelerator Facilities}
\vfill
\Author{Bhaskar Dutta\support}
\Address{\tamu}
\vfill
\begin{Abstract}
About 80 percent of the matter content of the universe is dark matter.
However, the particle origin of dark matter is yet to be established.
Many extensions of the Standard Model (SM) contain candidates of dark
matter. The search for the particle origin is currently ongoing at the large
hadron collider (LHC). In this review, I will summarize the different search
strategies for this elusive particle.
\end{Abstract}
\vfill
\begin{Presented}
Symposium on Cosmology and Particle Astrophysics 2013\\
Honolulu, HI,  November 12- 15, 2013
\end{Presented}
\vfill
\end{titlepage}
\def\thefootnote{\fnsymbol{footnote}}
\setcounter{footnote}{0}

\section{Introduction}

So far, the Higgs Boson is discovered at the LHC. However no new particle from any extension of the SM has been observed.  Interestingly, the mass of the Higgs Boson lies in the tight minimal supersymmetry (SUSY) Model's prediction window. This  serves as a tremendous motivation to investigate SUSY theories at the LHC. In SUSY theories, the Higgs divergence problem is resolved, grand unification of the gauge couplings can be achieved and the electroweak symmetry can be broken radiatively and a dark matter (DM) candidate can be obtained in supersymmetric SM which  can explain the precisely measured dark matter content of the universe.

 SUSY particles should  be directly observed at the large hadron collider.  In most R parity conserving SUSY models, the lightest supesymmetry particle (LSP) is $\tilde\chi^0_1$ is the DM candidate. The annihilation cross-section mostly involves sleptons, chargino, neutralinos etc. or SUSY particles without any color charge. These particles are light in most of the models compared to the colored particles. Producing and observing these particles are not easy at the LHC which mostly produces colored particles. The existence of the R parity conserving SUSY models will show up in  jets+$\met$, jets+leptons+$\met$ final states. The determination of the dark matter content and establshing SUSY or any new model will require establishing the predicted particle spectrum.

One way  to produce the lightest SUSY particle is via cascade decays from  productions of   $\tilde q\tilde g$, $\tilde q \tilde q$, $\tilde g\tilde g$ etc.
The squarks and gluinos then decay  into  quarks, heavier neutralinos and  charginos. The heavier neutralino and charginos then decay into lightest neutralino ($\tilde\chi^0_1$) and Higgs, $Z$, $W$,  leptons ($e$. $\mu$ etc), $\tau$s etc. The final state typically contains multiple leptons plus multiple jets plus \met.  $\tilde\chi^0_1$ is the DM candidate. 
In order to find the signal beyond the background, the event selection is made with large amount of  missing
energy,  high $p_T$ jets, large numbers  of jets and leptons. In order to establish the existence of various particles appearing in the cascades and measure their masses we need to measure the end-points of various distributions~\cite{hinchliffe}, $M_{ll}$, $M_{\tau\tau}$, $M_{j l}$, $M_{j\tau}$, $M_{j\tau\tau}$, $M_{jW}$, $p_T$ of $e$, $\mu$ and $\tau$ etc~\cite{duttaend,Dutta:2010uk,Dutta:2011kp}. These observables involve the sparticles in the cascade decays, e.g., sleptons, charginos, neutralinos etc.  We can use these observables to determine the masses of SUSY particles. However, determining the masses requires removal of combinatoric backgrounds from the SUSY and SM processes. We have developed bi-event subtraction techniques to remove these backgrounds to identify particular decay chain and showed how to determine masses in various models. This technique can be used not only to establish SUSY models, but also any  model with new colored states.

The non-colored SUSY particles e.g., charginos, neutralinos and sleptons which enter into the DM content calculation needs to be probed directly. In this review I mention one technique  called  vector boson fusion (VBF)  which provides a very promising search avenue not only for particle with electroweak charges but also SUSY particles with compressed spectra. This technique can be used not only in SUSY models but also for new particles with electroweak charges in non-SUSY theories. VBF has been considered in the context of Higgs searches~\cite{dawson} and in supersymmetric searches, in the context of $R-$parity violating MSSM~\cite{Dutta:2012xe,Delannoy:2013ata,datta}. The VBF production is characterized by the presence of two energetic jets in the forward direction in opposite hemispheres. The central region is relatively free of hadronic activity, and provides a potential probe of the supersymmetric EW sector, with the SM and $t \overline{t}+$ jets background under control. The VBF search strategies  can be used any model with new electroweak charged particles. 

It is also possible to probe DM particle at the LHC when only one jet accompanies the DM particle which is described as monojet analysis. The DM particles can be pair produced along  with a jet from initial state radiation to give rise to this signal which can be used to study interaction between the DM particle and SM particles.  The monojet signal also can arise when DM is singly produced along with a jet. In this type of models, the large missing energy is associated with an energetic jet whose transverse momentum distribution has a Jacobian-like shape.  The monojet analys can distinguish different dark matter models.

Finally,  the light stop squark  pair production at the LHC can be used to probe the nature of $\tilde\chi^0_1$ which is crucial to determine the dark matter content arising in SUSY models. For example, when $\tilde\chi^0_1$ is Bino type then  the lightest stop decays $100\%$ into a top plus the lightest neutralino, where as, if  the lightest neutralino is a mixture of Bino and Higgsino (which is more suited to satisfy the thermal dark matter content), the lightest stop decays mostly into $(i)$ a top quark plus the second or third lightest neutralino, and $(ii)$ a bottom quark plus the lightest chargino. We get different final states from the stop pair production depeneding on the nature of $\tilde\chi^0_1$.

In this review, I will discuss the probe of DM in four different ways at the LHC (i) using cascade decays of colored particles, squarks and gluinos (ii) using VBF, (iii) using monojet and (iv) using direct stop productions at the LHC. The techniques developed in the first two sections can be applied to non-SUSY models as well.

\section{SUSY particles in Cascade decay chains}
Heavy colored object decay into lighter particles in cascades. In order to measure the masses of all these particles and model parameters, we must be able to fully or partially reconstruct their cascade decays from the particles which can be detected. However, reconstructions of these decays become very difficult because it is impossible to know which particles come from the cascade decay we wish to reconstruct. The  inclusion of particles which do not come from the cascade decay of interest is referred to as combinatoric background.

\subsection{Bi-Event Subtraction Technique} 
The combinatoric background can be removed  in some cases by powerful subtraction techniques. For instance, reconstructing $Z$ boson decays into lepton pairs is easy since the leptons are easy to detect in the collider setting, and their charges can easily be measured. To reconstruct the $Z$ boson from the leptons, we collect  Opposite-Sign, Same-Flavor (OSSF) lepton pairs and construct the dilepton invariant mass for each pair. To model the combinatoric background, a sample of  Opposite-Sign, Opposite-Flavor (OSOF) lepton pairs is selected as well.  Performing the OSSF$-$OSOF subtraction of the invariant mass distributions, $m_{\ell \ell}^{\rm OSSF-OSOF} = m_{e^+ e^- / \mu^+ \mu^-} - m_{e^{\pm} \mu^{\mp}}$, yields a distribution which shows a clear peak of the $Z$ boson mass.

However, such subtraction techniques are not available for jets, whose charges and flavors are not easily determined and we therefore invoke the Bi-Event Subtraction Technique (BEST)~\cite{Dutta:2011gs}. In this technique, the combinatoric background of jets is modeled by combining jet information from a different event (or bi-event). For instance, during the reconstruction of the $W$ boson decaying into two jets, a signal may be seen if a sample of jet pairs is collected for each event to construct the dijet invariant mass distribution, $m_{jj}^{\rm same}$. Here, the \textquotedblleft same\textquotedblright\ suggests that the jet pairs arise from the same event. Some of the jet pairs in the same event distribution may originate from a single $W$ boson decay in the events, while other jet pairs will be combinatoric background. By taking a jet pair where each jet comes from a {\it different} event, the bi-event distribution, $m_{jj}^{\rm bi}$, can be formed. This bi-event distribution will have no jet pairs which stem from a single $W$ boson. This bi-event distribution models the combinatoric background well.

To demonstrate this powerful technique, we generate LHC events using \pythia~\cite{pythia} , and perform a LHC detector simulation using \pgs~\cite{pgs}. First,  SM $t\bar{t}$ events with some $W$+jets background (both at $7~\tev$ centre of mass collision energy) are used where $W$+jets signal is the main source of background for finding the top quark. We generate these  events using \alpgen~\cite{alpgen} and applied  \pythia\ and \pgs. We select events for analysis with the following cuts~\cite{topMassCDF}: (i) Number of leptons, $N_\ell = 1$, where $P_T^{(\ell)} \ge 20~\gev$; (ii) Missing transverse energy, $\met \ge 20~\gev$; (iii) Number of jets, $N_j \ge 3$, where $P_T^{(j)} \ge 30~\gev$ and at least one jet has been tightly $b$-tagged~\cite{pgs}; (iv) Number of taus, $N_\tau = 0$ for taus with $P_T^{(\tau)} \ge 20~\gev$~\cite{pgs}.

\begin{figure} 
\centering	
\includegraphics[width = 7.5cm]{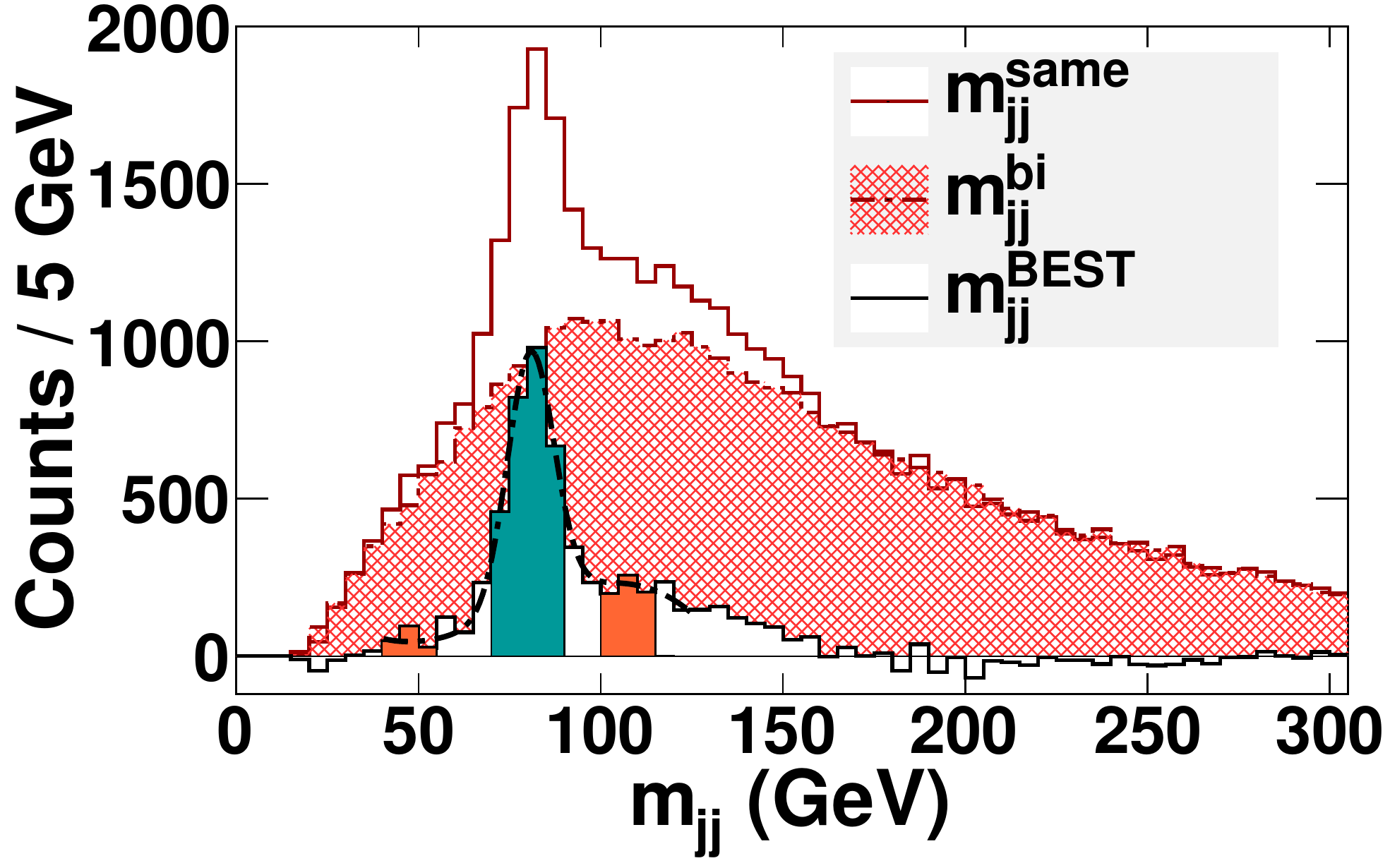}
	\caption{The dijet invariant mass distribution, $m_{jj}$. This plot shows the same-event ($m_{jj}^{\rm same}$), bi-event ($m_{jj}^{\rm bi}$), and BEST. For an integrated luminosity of $2~\invfb$, we show the $W$ boson mass, $m_W = 80.8 \pm 6.5~\gev$~\cite{Dutta:2011gs}.}
	\label{figTTbarBESTfindsW}
\end{figure}

With our selected events we pair up jets (which are not $b$-tagged) to fill the same-event and bi-event $m_{jj}$ distributions as described above. To fill the bi-event distribution, we refer to jets from the previous event. We perform  this for all of our events, where the $t\bar{t}$ and $W$+jets events have been mixed up randomly according to their production cross-sections. Once the distributions are filled, we normalize the shape of the $m_{jj}^{\rm bi}$ distribution to that of the $m_{jj}^{\rm same}$ distribution in the region $150~\gev < m_{jj} < 500~\gev$. Then we perform  BEST. The result of this subtraction is shown in Fig.~\ref{figTTbarBESTfindsW}

\begin{figure} 
\centering	
	\includegraphics[width = 7.5cm]{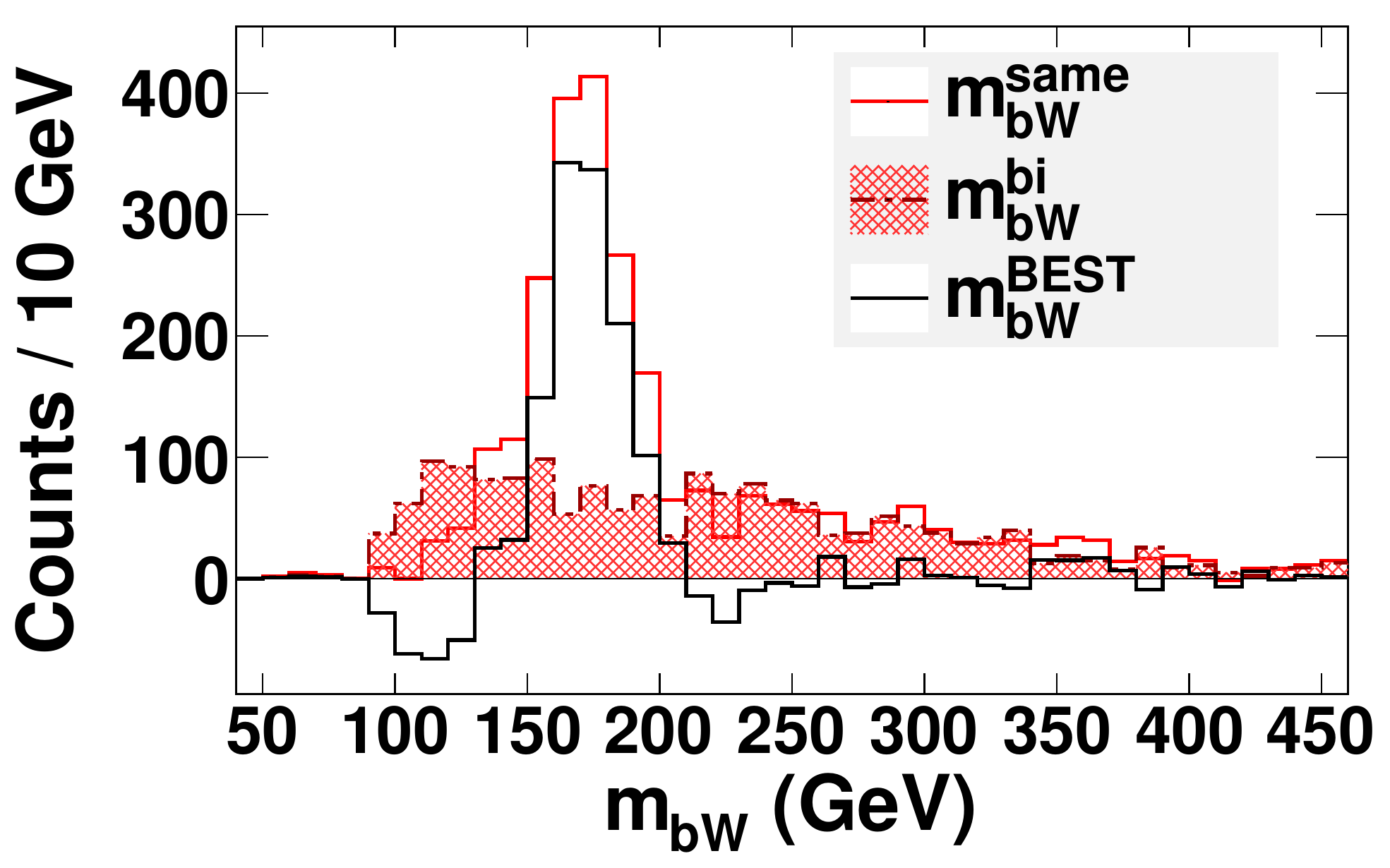}
	\caption{The $W$ plus $b$ invariant mass distribution, $m_{bW}$. This plot shows the same-event, bi-event, and BEST distributions. For an integrated luminosity of $2~\invfb$, we show the top quark mass, $m_t = 172 \pm 13~\gev$. The top quark mass is set within \alpgen\ as $m_t = 174.3~\gev$~\cite{Dutta:2010uk}}
	\label{figTTbarBESTfindsTop}
\end{figure} 

\vspace{0.2cm}
Once we have found the $W$ boson with this first application of BEST, we can combine the $W$ boson with a $b$-jet to find the top quark. To remove additional background from the $W$ signal, we perform a sideband subtraction.

Finally in order to remove the combinatoric background from $b$-jets not from the same top quark as the $W$ boson, we apply BEST again which models the combinatoric background very well, since the $W$ and $b$ from different events cannot possibly come from a single top quark.  The resulting $m_{bW}$ distribution is shown in Fig.~\ref{figTTbarBESTfindsTop}. 

We have  used BEST to determine masses, model parameters. We usee the masse and model parameters to estimate the DM content in non-universal supergravity (nuSUGRA)~\cite{Dutta:2010uk} and mirage mediation models~\cite{Dutta:2011kp}. In this review talk, I will describe the case of nuSUGRA model only.

\subsection{Non-Universal SUGRA model}
We first review the minimal supergravity (mSUGRA)/Constrained MSSM (CMESSM) model~\cite{sugra1} in order to describe the nuSUGRA model. The mSUGRA model has the attractive feature that many of the SUSY particle masses are unified at the GUT scale which means that it needs only four parameters and a sign to specify the entire model. These parameters are:
\begin{itemize}
\item The unified scalar mass at the GUT scale, $\mzero$,
\item The unified gaugino mass at the GUT scale, $\mhalf$,
\item The trilinear coupling at the GUT scale, $\azero$,
\item The ratio of the vacuum expectation values of the two Higgs
doublets, $\tanb$, and
\item The sign of the Higgs bilinear coupling, $\mathrm{sign}(\mu)$.
\end{itemize}

 In the nuSUGRA model, the Higgs bosons are given a non-universal mass. 
Since there are two Higgs doublets in SUSY models, we can have a parameter for each of their masses at the GUT scale, i.e., $m_{H_u}^2=(1+\delta_{H_u}) m_0^2$, $m_{H_d}^2=(1+\delta_{H_d}) m_0^2$. However, only one of the Higgs masses affects the  parameter $\mu$ which becomes a free parameter in this model. The value of $\mu^2$ at the electroweak scale in terms of the GUT scale parameters is determined by the renormalization group equations (RGEs). In general, one must solve these numerically. However, one can get a qualitative understanding of the effects of the $\delta_H$'s from an analytic solution which is valid for low and intermediate $\tanb$~\cite{nath1}: 
\begin{equation} 
\mu^2=\frac{t^2}{t^2-1} \left[
	\left(\frac{1-3 D_0}{2}-\frac{1}{t^2}\right)
	+\left(-\frac{1+D_0}{2}\delta_{H_u}+\frac{\delta_{H_d}}{t^2}\right)
\right]m_0^2 + \Delta,
\label{eqMu}
\end{equation} 
where $t\equiv \tanb$, $D_0\simeq 1-(m_t/200\ \sinb)^2$, and $\Delta$ contains the universal parts (which are independent of the $\delta_H$'s) and loop corrections. In general $D_0$ is small ($D_0\leq 0.23$). Equation \ref{eqMu} shows that $\mu$ is primarily sensitive to $\delta_{H_u}$. However, the pseudoscalar and heavy Higgs boson masses depend on both $\delta_{H_u}$ and $\delta_{H_d}$. 

In this model, the DM content can be satisfied not only by lowering $\mu$ but also having the pseuoscalar or heavy Higgs mass equal to twice the neutralino mass. Since we have two new parameters in the Higgs sector, both the pseudoscalar mass and $\mu$ are free parameters in this model. In the case where the DM content is satisfied by the heavy Higgs/pseudoscalar Higgs resonance, the heavy Higgs mass needs to be measured to see whether its mass obeys the resonant funnel condition. In our case we do not consider the Higgs funnel region but consider the first scenario where $\mu$ is changed to satisfy the DM content. For the purposes of this study, we choose one such model which predicted a DM relic density in agreement with that measured by WMAP. This scenario is also quite interesting since it has large direct detection spin-independent cross-section, and therefore it will be detected in the ongoing/upcoming runs of direct detection experiments.

Since $\mu$ is affected by only the up type Higgs, we define the nuSUGRA model with the unified Higgs mass at the GUT scale, $m_{H_u} = m_{H_d} \equiv m_H$, which becomes the fifth parameter of the model. The mass spectrum for our benchmark point of the nuSUGRA model is shown in Table~\ref{tabSpectrum} where the mass spectrum for this model is determined using \isasugra~\cite{isajet}. 

\begin{table}
\caption{SUSY masses and parameters (in $\gev$) for the point $\mzero = 360~\gev$,
$\mhalf = 500~\gev$, $\tanb = 40$, $\azero = 0$, and $m_H = 732~\gev$.  The top mass is set as $172.6~\gev$.
For this point, the DM relic density is $\DMrelic = 0.11$. The
total production cross-section for this point is $\sigma =
1.25~\mathrm{pb}$.}
 \label{tabSpectrum}
\begin{tabular}{c c c c c c c c c c | c}
\hline \hline
$\gluino$ &
$\begin{array}{c} \usquarkL \\ \usquarkR \end{array}$ &
$\begin{array}{c} \stoptwo \\ \stopone \end{array}$ &
$\begin{array}{c} \sbottomtwo \\ \sbottomone \end{array}$ &
$\begin{array}{c} \seleL \\ \seleR \end{array}$ &
$\begin{array}{c} \stautwo \\ \stauone \end{array}$ &
$\begin{array}{c} \schitwozero \\ \schionezero \end{array}$ &
$\begin{array}{c} \schifourzero \\ \schithreezero \end{array}$ &
$\begin{array}{c} \schitwopm \\ \schionepm \end{array}$ &
$\begin{array}{c} A^0 \\ h^0 \end{array}$ &
$\mu$
\\ \hline
1161 &
$\begin{array}{c} 1113 \\ 1078 \end{array}$ &
$\begin{array}{c} 992 \\ 781 \end{array}$ &
$\begin{array}{c} 989 \\ 946 \end{array}$ &
$\begin{array}{c} 494 \\ 407 \end{array}$ &
$\begin{array}{c} 446 \\ 255 \end{array}$ &
$\begin{array}{c} 293 \\ 199 \end{array}$ &
$\begin{array}{c} 432 \\ 316 \end{array}$ &
$\begin{array}{c} 427 \\ 291 \end{array}$ &
$\begin{array}{c} 647 \\ 115 \end{array}$ &
307 
\\ \hline \hline
\end{tabular}
\end{table}

\subsubsection{Characteristic Signal and Observables at the LHC}

Our benchmark point of the nuSUGRA model shows  that the LHC would see high $\pt$ jets from squark decays to neutralinos and charginos, many $\tau$'s from neutralino and stau decays, and large missing transverse energy ($\met$) from the lightest neutralino escaping the detector. There are also many $W$ bosons being produced from neutralinos decaying into charginos or vice versa.
The mass spectrum is processed by \pythia. These events are $14~\tev$ $pp$ collisions which are  then passed on to \pgs to simulate the detector effects. 

Since we require five independent measurements to determine our five model parameters, we construct as many useful measurements as possible. In that process we have found that it is necessary to utilize the $W$ boson decay chains.

\subsubsection{Determining Model Parameters and Relic Density}

We reconstruct the following six observables, $\meffpeak$, $\meffbnoWpeak$, $\mjWend$, $\mjttpeak$, $\mttend$, $\mjtend$ and  we determine the model parameters using them. 

\begin{table}
\caption{Results from the fits of kinematical observables found at our benchmark point, along with its statistical uncertainty for luminosities of $1000~\invfb$ and $100~\invfb$, and its systematic uncertainty. All values have units of $\gev$.}
\label{tabObservables}
\begin{tabular}{c | c c c c}
\hline \hline
Observable & Value & $1000~\invfb$ Stat. & $100~\invfb$ Stat. & Systematic \\ \hline
$\meffpeak$ & 1499 & $\pm 7$ & $\pm 21$ & $\pm 45$\\
$\meffbnoWpeak$ & 1443 & $\pm 43$ & $\pm 107$ & $\pm 43$\\ 
$\mjWend$ & 793 & $\pm 2$ & $\pm 5$ & $\pm 29$\\
$\mjttpeak$ & 415 & $\pm 8$ & $\pm 26$ & $\pm 40$ \\ 
$\mttend$ & 85.3 & $\pm 0.8$ & $\pm 2.8$ & $\pm 3.8$ \\
$\mjtend$ & 540 & $\pm 2$ & $\pm 6$ & $\pm 34$ \\ \hline \hline
\end{tabular}
\end{table}

Once we have finally determined all the model parameters, we use \darksusy~\cite{darksusy} to calculate the DM relic density of the universe today, $\DMrelic$. We also estimate the uncertainty in the DM relic density due to the uncertainties in the measured model parameters. Our results are shown in Table~\ref{tabResults}. We find that the model parameters $\mzero$, $\mhalf$, $m_H$ and $\tanb$ can be determined a good accuracy: The statistical uncertainties are $\leq 15\%$ for $100~\invfb$ luminosity, with the systematic uncertainties nearly the same. 

We can determine the accuracy of $\mu$ from these parameters and we find that $\mu$ can be determined with accuracies of around 15\% and 8\% for luminosities of $100~\invfb$ and $1000~\invfb$, respectively. The uncertainty of $\mu$ is influenced not only by the uncertainty in $m_H$, but by $\mzero$ and other model parameters as well, as obtained from Equation~\ref{eqMu}.

\begin{figure} [t!]
\centering
\includegraphics[width=.50\textwidth]{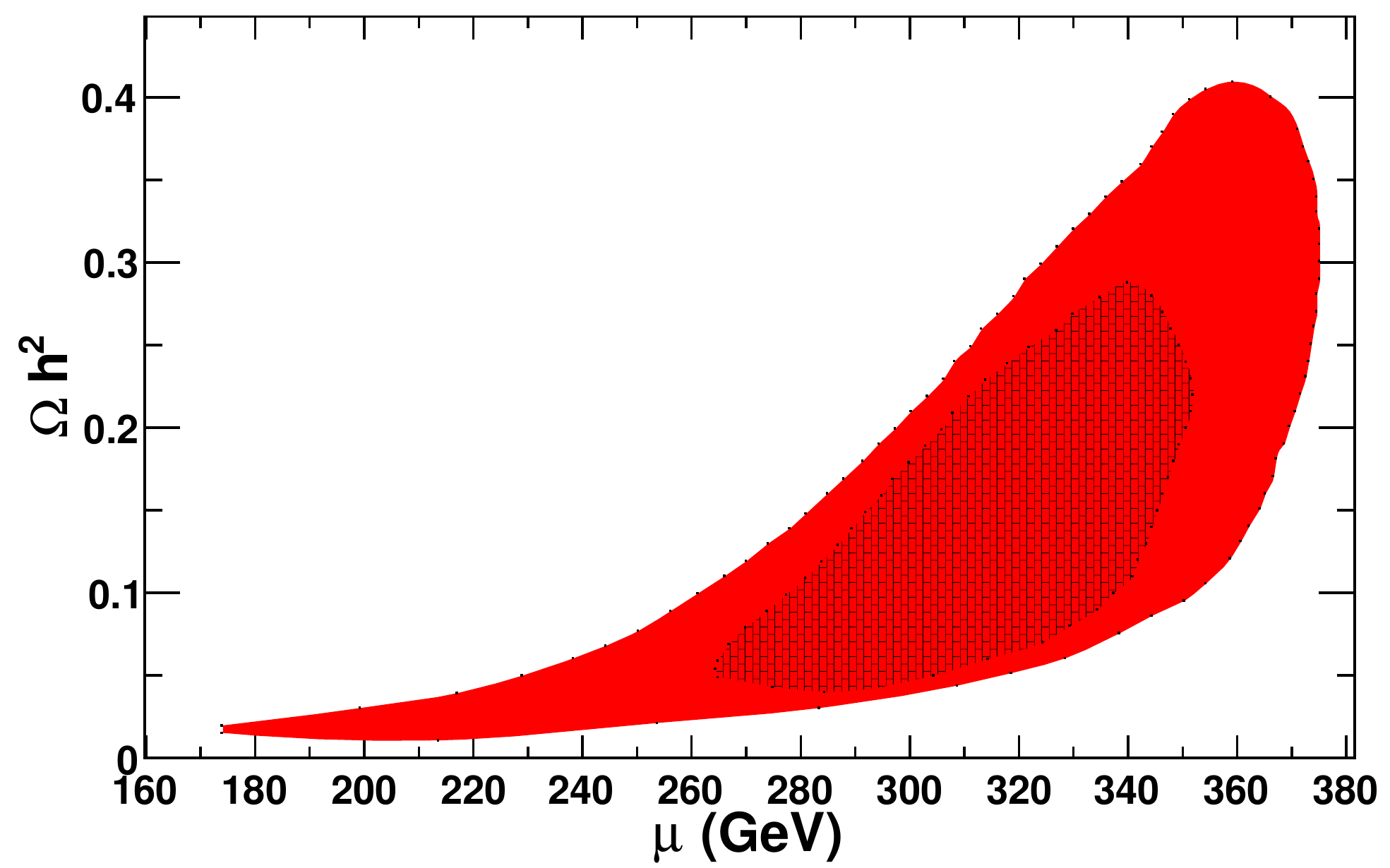}
\caption{Estimates of the statistical $1\sigma$ uncertainties in the $\DMrelic$ versus $\mu$ plane. The solid red (brick textured) region is for a luminosity of $100~\invfb$ ($1000~\invfb$)~\cite{Dutta:2010uk}.}
\label{figEllipse}
\end{figure}

Since the DM content is sensitive to the value of $\mu$, in Fig.~\ref{figEllipse}, we plot one $\sigma$ contours of the DM content as a function of $\mu$ for luminosities of $100~\invfb$ (red shaded region) and $1000~\invfb$ (brick shaded region). The determination of DM content is of couse much better with $1000~\invfb$, but even with $100~\invfb$ the measurement accuracy is quite encouraging.

\begin{table}
\caption{Results of the nuSUGRA model parameters and relic density of DM in the universe for integrated luminosities of $1000~\invfb$ and $100~\invfb$. The systematic uncertainties are also estimated here. Note that the uncertainties for an integrated luminosity of $100~\invfb$ were estimated by simply scaling down the distributions before performing fits for the analysis.}
 \label{tabResults}
\begin{tabular}{c | c c c c c | c c}
\hline \hline
$\lum~(\invfb)$ & $\mhalf$ & $m_H$ & $\mzero$ &
$\azero$ & $\tanb$ & $\mu$ & $\DMrelic$ \\ \hline
$1000$ & $500 \pm 3$ & $727 \pm 10$ & $366 \pm 26$ & 
$3 \pm 34$ & $39.5 \pm 3.8$ & $321 \pm 25$ & $0.094^{+0.107}_{-0.038}$ \\ 
$100$ & $500 \pm 9$ & $727 \pm 13$ & $367 \pm 57$ & 
$0 \pm 73$ & $39.5 \pm 4.6$ & $331 \pm 48$ & $0.088^{+0.168}_{-0.072}$ \\ 
Syst. & $\pm 10$ & $\pm 15$ & $\pm 56$ & 
$\pm 66$ & $\pm 4.5$ & $\pm 48$ & $ ^{+0.175}_{-0.072}$ \\ \hline \hline
\end{tabular}
\end{table}

\section{Productions of DM particles via VBF}

Vector boson fusion (VBF) processes, characterized by two jets with large dijet invariant mass in the forward region in opposite hemispheres, are a promising avenue to search for new physics. Two recent studies have used VBF processes to investigate the chargino/neutralino sector of supersymmetric theories (\cite{Delannoy:2013ata}, \cite{Dutta:2012xe}). In \cite{Delannoy:2013ata}, direct DM production by VBF processes in events with $2j \, + \, \met$ in the final state was studied at the 14 TeV LHC, providing a search strategy that is free from trigger bias. Information about production cross sections in VBF processes and the distribution of $\met$ in the final state was used to solve for the mass and composition of the lightest neutralino, and hence the DM relic density. In \cite{Dutta:2012xe}, the second lightest neutralino and the lightest chargino were probed using VBF processes, in the $2j + 2\mu + \met$ (light slepton case) and  $2j + 2\tau + \met$ (light stau case) final states  at 8 TeV LHC. 


\subsection{Productions of Charginos, Neutralinos via VBF}
\begin{figure}[!htp]
\centering
\includegraphics[width=2.0in]{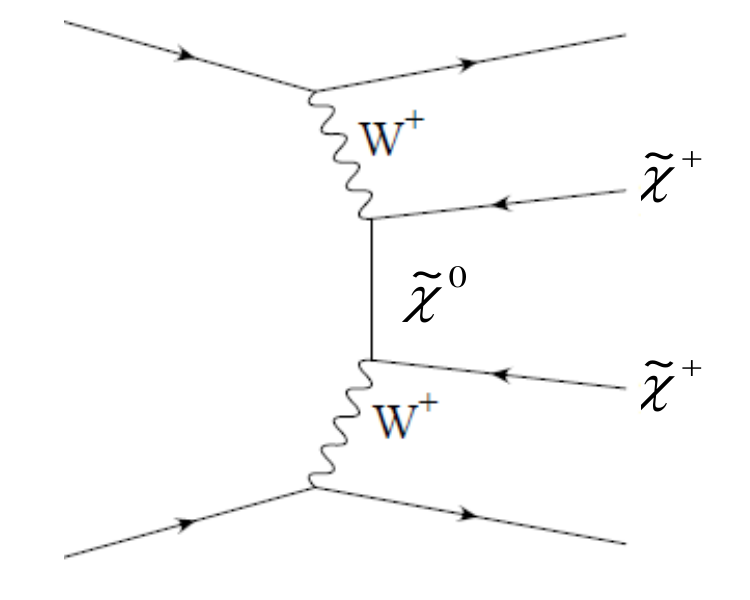}
\caption{Charginos pair productions by VBF processes are presented~\cite{Dutta:2012xe}.}
\label{VBFCharginoDiagram}
\end{figure}

  A sample production of chargino  from VBF processes is shown in Figure \ref{VBFCharginoDiagram}.

For $R-$parity conserving models, the decay of the lightest chargino and second lightest neutralino can be written as
\bea
\chpm{1} \, \rightarrow \, \tilde{\tau}^{\pm}_1 \nu \, \rightarrow \, \tau^{\pm} \neuo{1} \nu \nonumber \\
\neuo{2}  \, \rightarrow \, \tilde{\tau}^{\pm}_1 \tau^{\mp} \, \rightarrow \, \tau^{\pm} \tau^{\mp} \neuo{1} \,\,.
\eea
The staus are assumed to be lighter than the lightest charginos and second lightest neutralino. If   $\tilde e$/$\tilde\mu$ are lighter than the lightest charginos and second lightest neutralino then  $e$s/$\mu$s will be present in the final state. This  scenario will be discussed as well. 

The production of  $VV$ (where $V$ is $W$ or $Z$) by VBF processes are backgrounds to the signal states. A $\met$ cut is needed to reduce this background. 
Moreover, requiring multiple $\tau$'s in the event further reduces background and the results are shown  requiring same-sign and oppositely-signed $\tau$ pairs, as well as an inclusive study. 
%
%
Although  $m_{\tilde{\chi}^{\pm}_1} \, \sim \, m_{\tilde{\chi}^0_2}$ is chosen as an example, the methods described here can be  applicable in detecting $\neuo{2}$ and $\chpm{1}$ separately.

The VBF selections  are~\cite{Dutta:2012xe} 

$(i)$ The leading jet has $p_T \geq 75$ GeV, all other jets have $p_T \geq 50$ GeV in $|\eta| \leq 5$;

$(ii)$ $|\Delta \eta (j_1,j_2)| > 4.2$, where $j_1$ and $j_2$ are any jets with $p_T \geq 50$ GeV in $|\eta| \leq 5$;

$(iii)$ $\eta_{j_1} \eta_{j_2} < 0$;

$(iv)$ $M_{j_1j_2} > 650$ GeV, where $M_{j_1j_2}$ is the largest dijet invariant mass of all possible jet pairs. Using all possible jet pairs is less sensitive  on the signal acceptance due to initial/final state radiation, pileup, and fluctuations in jet fragmentation.

\begin{figure}[!htp]
\centering
\includegraphics[width=3.5in]{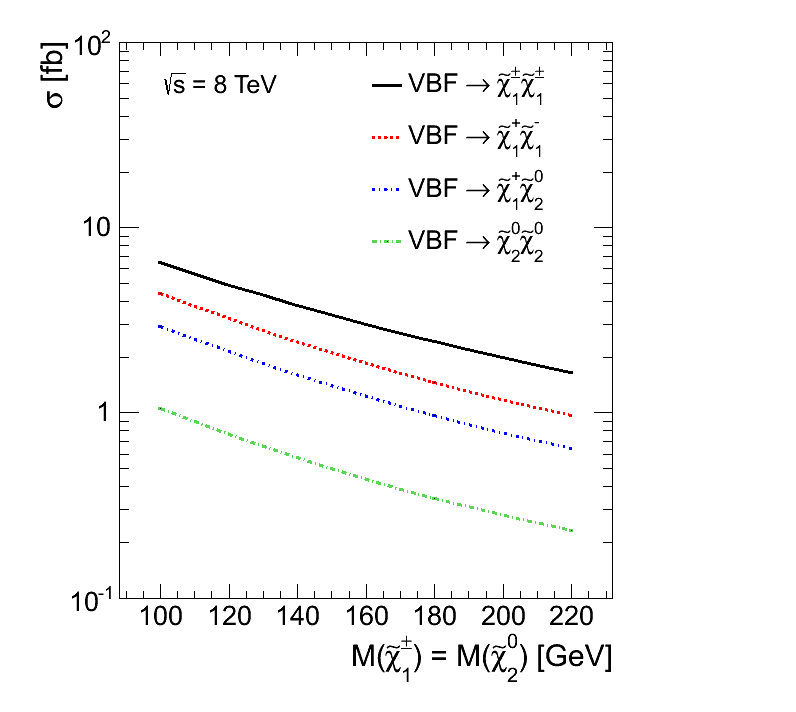}
\caption{VBF production cross-section at $\sqrt{s} = 8$ TeV are presented as a function of mass for various channels after imposing $|\Delta \eta| \, > \, 4.2$~\cite{Dutta:2012xe} .}
\label{VBFXsection}
\end{figure}

The production cross-sections are shown at $\sqrt{s} = 8$ TeV for $\chpm{1} \chpm{1}, \chip{1} \chim{1}, \chpm{1} \neuo{2},$ and $\neuo{2} \, \neuo{2}$ as a function of mass after imposing just $|\Delta \eta| \, > \, 4.2$  in Figure \ref{VBFXsection}. 


The large production cross-sections for $\chpm{1} \chpm{1}$ and $\chpm{1} \neuo{2}$ mean that same-sign $\tau$'s are significantly present  and the same-sign $\tau$ selection leads to considerable reduction of background.


\begin{table}[!htp] 
\caption{The  cross section (fb) for inclusive $\chpm{1} \chpm{1}, \chip{1} \chim{1}, \chip{1}\neuo{2}$ and $\neuo{2} \neuo{2}$ pair production by VBF processes and main backgrounds are given in the $\geq 2j + 2\tau$ final state at $\sqrt{s} = 8$ TeV. Results for same-sign and oppositely-signed final state $\tau$ pair, as well as inclusive study, are shown. All masses  are in GeV. The significance is shown for $25$ fb$^{-1}$.}
\label{tabletautau}
\begin{tabular}{c c c c c c} 
\hline \hline 
                    

                               &Signal    &\,\, $Z+$jets &\,\, $W+$jets &\,\, $WW$ &\,\, $WZ$ \\
          
\hline 
              \hline  \\
                    



                     VBF cuts            &$4.61$   & $10.9$                        &$3.70 \times 10^3$                      &$97.0$  &$19.0$      \\      
      $\met > 75$                 &$4.33$   & $0.27$                        &$5.29 \times 10^2$                      &$17.6$  &$3.45$      \\
      \hline
       $2 \, \tau,$ inclusive                  & $0.45$       & $0.06$                             &$0.23$                           &$0.09$       &$0.04$          \\   
        $(S/\sqrt{B})$                       & \multicolumn{5}{c}{$3.47$} \\ 
       \hline
      $\tau^{\pm} \tau^{\pm}$                  &$0.21$       &$0$                             &$0.11$                           &$0.02$       &$0.01$          \\                          
    $(S/\sqrt{B})$                       & \multicolumn{5}{c}{$2.91$} \\         
\hline
      $\tau^{\pm} \tau^{\mp}$                  &$0.24$       &$0.06$                             &$0.12$                           &$0.07$       &$0.03$          \\   
     $(S/\sqrt{B})$                       & \multicolumn{5}{c}{$2.27$} \\       
\hline \hline

\end{tabular}
\end{table}

A summary of the effective cross-section at each selection stage of the study is given in Table \ref{tabletautau} for the main sources of background and inclusive $\chpm{1} \chpm{1}, \chip{1} \chim{1}, \chpm{1}\neuo{2}$ and $\neuo{2} \neuo{2}$ pair production by VBF processes. The VBF and $\met$ cuts are very effective in reducing the background. The significance (at $25$ fb$^{-1}$) for inclusive, opposite-sign, and like-sign $\tau$ pairs are $3.47,\, 2.91,$ and $2.27$, respectively at the benchmark point, where $\Delta M \, = \, m_{\tilde{\tau}_1} - m_{\neuo{1}} \, = \, 30$ GeV. For $\Delta M \, = \, 15$ GeV, the significance (at $25$ fb$^{-1}$) for inclusive, opposite-sign, and like-sign $\tau$ pairs are $1.0,\,  0.83,$ and $0.66$, respectively. Although the search with $\tau$ leptons are  very hard, we also investigated the  sensitivity in cases where the lightest chargino and the second lightest neutralino  decay to the first two generation of sleptons.

 After VBF selections, the following selections are employed:

$(i)$ Two isolated $\mu$'s with $p_T \geq 20$ GeV and $p_T \geq 15$ GeV in $|\eta| < 2.1$.

$(ii)$ $\met > 75$ GeV.

For this signal, the VBF and $\met$ cuts are very effective in reducing the background. 
We find that the  significances (at $25$ fb$^{-1}$) for inclusive, opposite sign, and like-sign $\mu$ pairs are $13.5, \, 15.4,$ and $7.80$, respectively and the significance for opposite-sign $\mu$ pairs drops to $\sim 3.0$ for $m_{\tilde{\chi}^{\pm}_1} \, \sim \, m_{\tilde{\chi}^0_2} \, \sim 330$ GeV.

\subsection{Production the Lightest Neutralinos Directly via VBF}
In order to probe DM directly, we investigate the following processes~\cite{Delannoy:2013ata}:
%
%
\be
pp \rightarrow \neuo{1} \, \neuo{1} \, jj, \,\,\, \chpm{1} \, \chmp{1} \, jj , \,\,\, \chpm{1} \, \neuo{1} \, jj \,\,\,\,.
\end{equation}
The main sources of SM background for the study of dark matter using VBF production are: $(i)$ \,\, $pp \rightarrow Z jj \rightarrow \nu \nu j j $ and $(ii)$ \,\, $pp \rightarrow W jj \rightarrow l \nu j j $. The former is an irreducible background with the similar topology as the signal where the $\met$ arises from the neutrinos. The latter appears from events which survive a lepton veto; $(iii)$ \,\, $pp \rightarrow t\tbar + $jets which can be removed by vetoing $b$-jets, light leptons, $\tau$ leptons and light-quark/gluon jets.

The search strategy is based upon on requiring the tagged VBF jets, vetoes for $b$-jets, light leptons, $\tau$ leptons and light-quark/gluon jets, and requiring large $\met$ in the event. Like before, we generate the signal and background events with \madgraph\, \cite{Alwall:2011uj}. The \madgraph \, events are then passed through \pythia\,\cite{pythia} for parton showering and hadronization. The detector simulation code used here  is \pgs \, \cite{pgs}. 

The distributions of $\pt(j_1), \pt(j_2), M_{j_1j_2}$, and $\met$ for background as well as VBF pair production of DM at $\sqrt{s}=8$ TeV and 14 TeV are studied. For pure Wino or Higgsino DM, $\chpm{1}$ is taken to be outside the exclusion limits for ATLAS' disappearing track analysis \cite{chargedtrack} and thus VBF production of $\chpm{1} \chpm{1}$, $\chpm{1} \chmp{1}$, and $\chpm{1} \neuo{1}$ also contribute. The $\neuo{1}$ masses chosen for this study to be in the range 100 GeV to 1 TeV. The colored sector is assumed to be much heavier and therefore provides no contribution to the neutralino production from cascade decays of colored particles. 

We first preselect the events using  $\met > 50$ GeV and the two leading jets ($j_{1}$,$j_{2}$) where each jet each satisfying $p_{T} \geq 30$ GeV with $|\Delta \eta(j_{1},j_{2})| > 4.2$ and $\eta_{j_{1}}\eta_{j_{2}} < 0$ and the preselected events are then used to optimize the final selections to achieve maximal signal significance ($S / \sqrt{S + B}$). We employ the following cuts for the final selection: $(i)$ the tagged jets are required to have $p_{T} > 50$ GeV and $M_{j_{1} j_{2}} > 1500$ GeV; $(ii)$ events with loosely identified leptons ($l = e,\mu,\tau_{h}$) and $b$-quark jets are rejected which help to reduce the $t \tbar$ and $Wjj \rightarrow l\nu jj$ backgrounds by approximately $10^{-2}$ and $10^{-1}$, respectively, while achieving $99\%$ efficiency for signal events. The $b$-jet tagging efficiency  is $70\%$ with a misidentification probability of $1.5\%$, following Ref. \cite{Chatrchyan:2012jua}. We reject events with a third jet (with $p_{T} > 50$ GeV) residing between $\eta_{j_{1}}$ and $\eta_{j_{2}}$; $(iii)$ We optimize $\met$ cut for each different value of the DM mass. For $m_{\neuo{1}} = 100$ GeV ($1$ TeV), $\met \geq 200$ GeV ($450$ GeV) is chosen, reducing the $Wjj\rightarrow l\nu jj$ background by approximately $10^{-3} \, (10^{-4})$. 
We have found that missing energy is the biggest discriminator between background and signal events. After the missing energy cut, the azimuthal angle difference of the two tagging jets~\cite{zeppenfeld} does not improve the search limit.

\begin{figure}[!htp]
\centering
\includegraphics[width=3in]{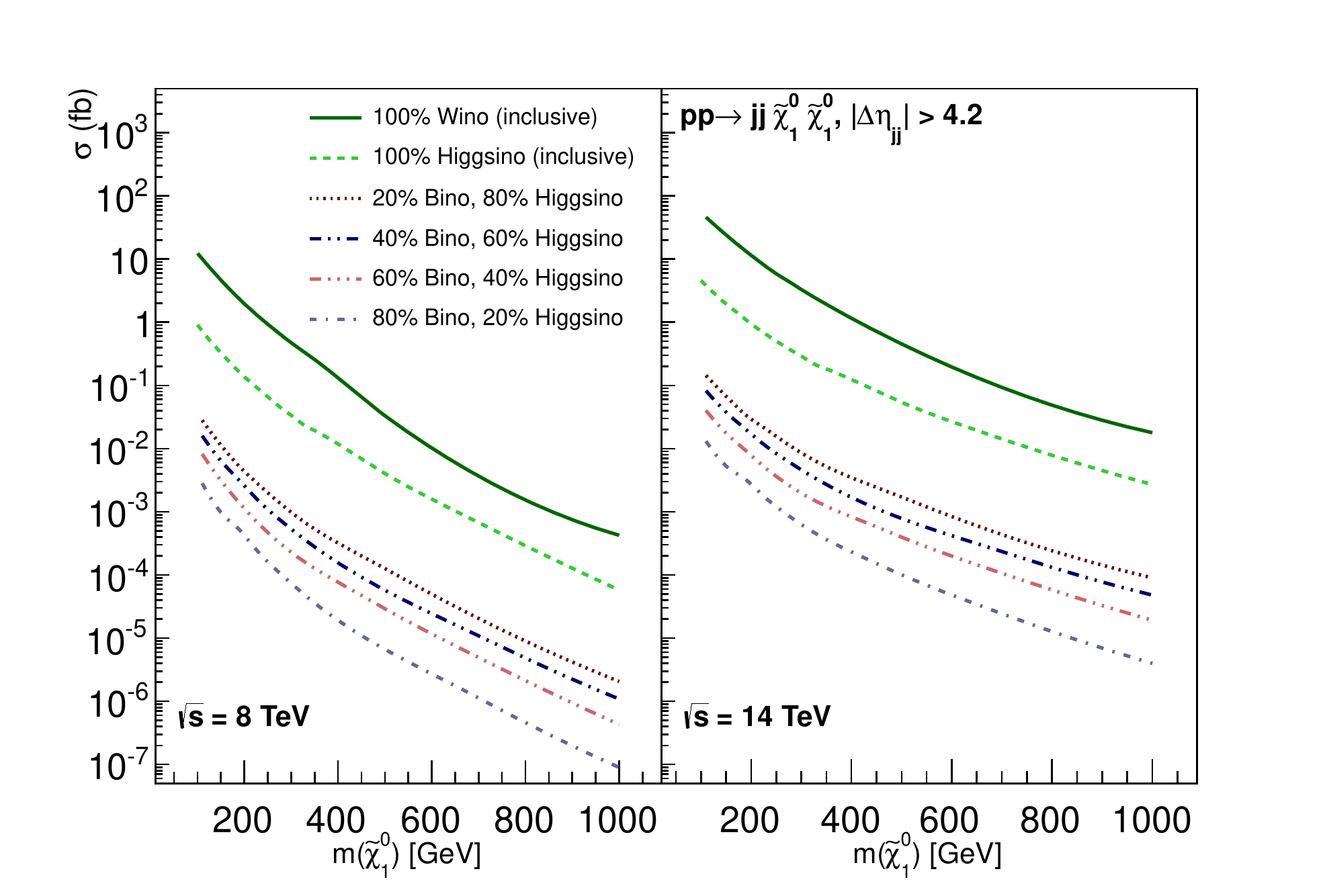}
\caption{Production cross section as a function of $m_{\neuo{1}}$ after requiring $|\Delta \eta (j_1, j_2)| > 4.2$, at LHC8 and LHC14. For the pure Wino and Higgsino cases, inclusive $\neuo{1} \neuo{1}$,  $\chpm{1} \chpm{1}$, $\chpm{1} \chmp{1}$, and $\chpm{1} \neuo{1}$ production cross sections are shown~\cite{Delannoy:2013ata}.}
\label{xsectionColor}
\end{figure}

In Fig.\ref{xsectionColor}, we show the production cross section is shown as a function of $m_{\neuo{1}}$ after requiring $|\Delta \eta (j_1, j_2)| > 4.2$. The left and right panels show the production cross-sections of lightest neutralinos for LHC8 and LHC14, respectively. For the pure Wino and Higgsino cases, inclusive $\neuo{1} \neuo{1}$,  $\chpm{1} \chpm{1}$, $\chpm{1} \chmp{1}$, and $\chpm{1} \neuo{1}$ production cross sections are shown. The green (solid) curve corresponds to the case where $\neuo{1}$ is $99\%$ Wino. The inclusive production cross section is $\sim 40$ fb for a $100$ GeV Wino at LHC14, and falls  with increasing mass. The cross section is  $\sim 5-10$ times smaller in the pure Higgsino case, represented by the green (dashed) curve. As the Higgsino fraction in $\neuo{1}$ decreases and Bino fraction increases for a given mass, the cross section drops. For $20\%$ Higgsino fraction and $80\%$ Bino in $\neuo{1}$, the cross section is $ \sim 10^{-2}$ fb for $m_{\neuo{1}} = 100$ GeV at LHC14.

\begin{figure}[!htp]
\centering
\includegraphics[width=3.5in]{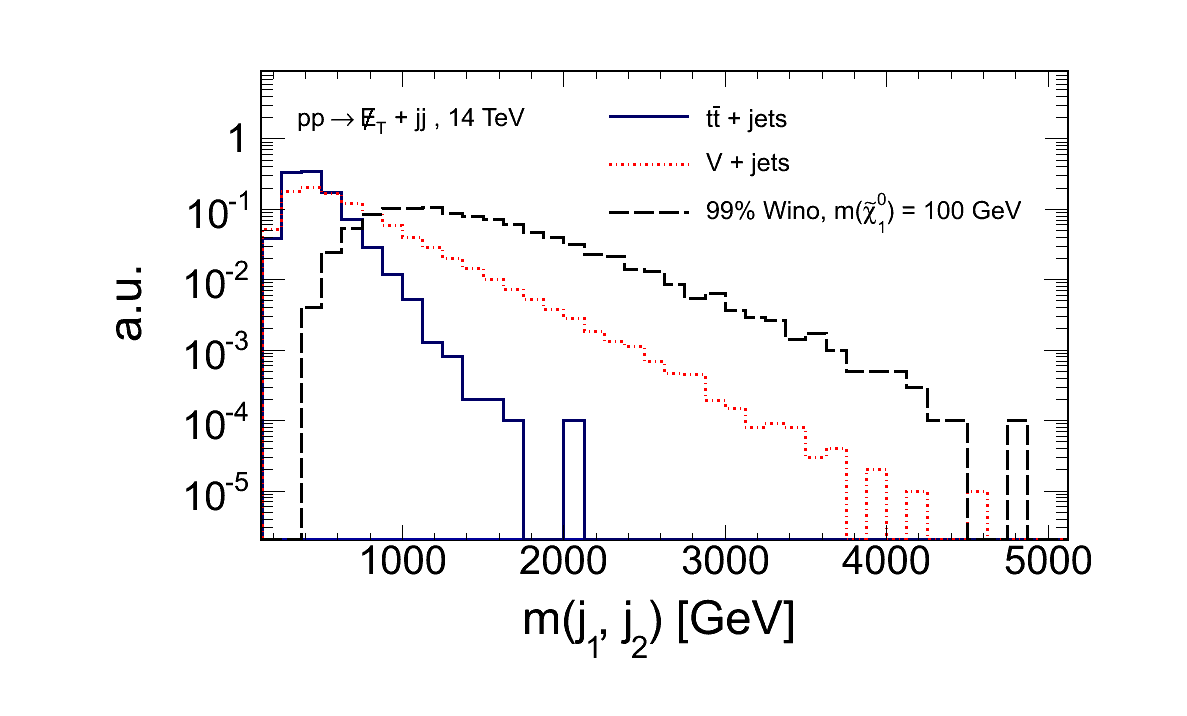}
\caption{We show the distribution of the dijet invariant mass $M_{j_1j_2}$ normalized to unity for the tagging jet pair $(j_1,j_2)$ and the main sources of background after pre-selection cuts and requiring $p_T > 50$ GeV for the tagging jets at LHC14. The dashed black curves display the distribution for the case where $\neuo{1}$ is a nearly pure Wino with $m_{\neuo{1}} = 50$ and $100$ GeV. Inclusive $\neuo{1} \neuo{1}$,  $\chpm{1} \chpm{1}$, $\chpm{1} \chmp{1}$, and $\chpm{1} \neuo{1}$ production is considered~\cite{Delannoy:2013ata}.}
\label{DiJetMass_VBFDM}
\end{figure}

We display the dijet invariant mass distribution $M_{j_1j_2}$ for the tagging jet pair $(j_1,j_2)$ and main sources of background in Figure \ref{DiJetMass_VBFDM} after the pre-selection cuts where we  require $p_T > 50$ GeV for the tagging jets at LHC14. The dashed black curves display the distribution for the case of a pure Wino type DM, with $m_{\neuo{1}} = 50$ and $100$ GeV.  The dijet invariant mass distribution for the SM background $W+$ jets, $Z+$ jets, and $t \bar{t} +$ jets are also displayed. Clearly, requiring $M_{j_1j_2} > 1500$ GeV is effective in rejecting background events, resulting in a reduction rate between $10^{-4}$ and $10^{-2}$ for the backgrounds of interest.

\begin{figure}[!htp]
\centering
\includegraphics[width=3.5in]{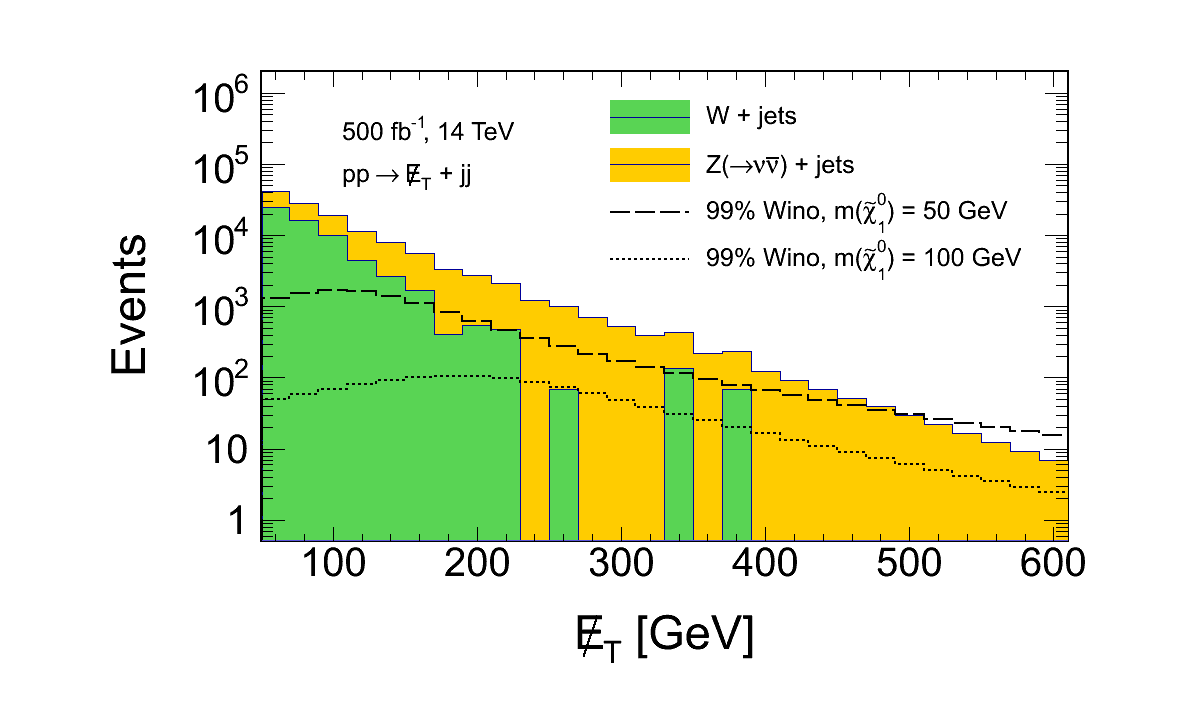}
\caption{The $\met$ distributions for Wino DM (50 GeV and 100 GeV) compared to $W+$ jets and $Z+$ jets events with $500$ fb$^{-1}$ integrated luminosity at LHC14. The distributions are after all selections except the $\met$ cut. Inclusive $\neuo{1} \neuo{1}$,  $\chpm{1} \chpm{1}$, $\chpm{1} \chmp{1}$, and $\chpm{1} \neuo{1}$ production is considered~\cite{Delannoy:2013ata}.}
\label{Met_VBFDM_versionB}
\end{figure}

We show the $\met$ distribution for an integrated luminosity of 500 fb$^{-1}$ at LHC14  after all final selections except the $\met$ requirement in Figure \ref{Met_VBFDM_versionB}. We find that there is a significant enhancement of signal events in the high $\met$ region.


\begin{figure}[!htp]
\centering
\includegraphics[width=3.5in]{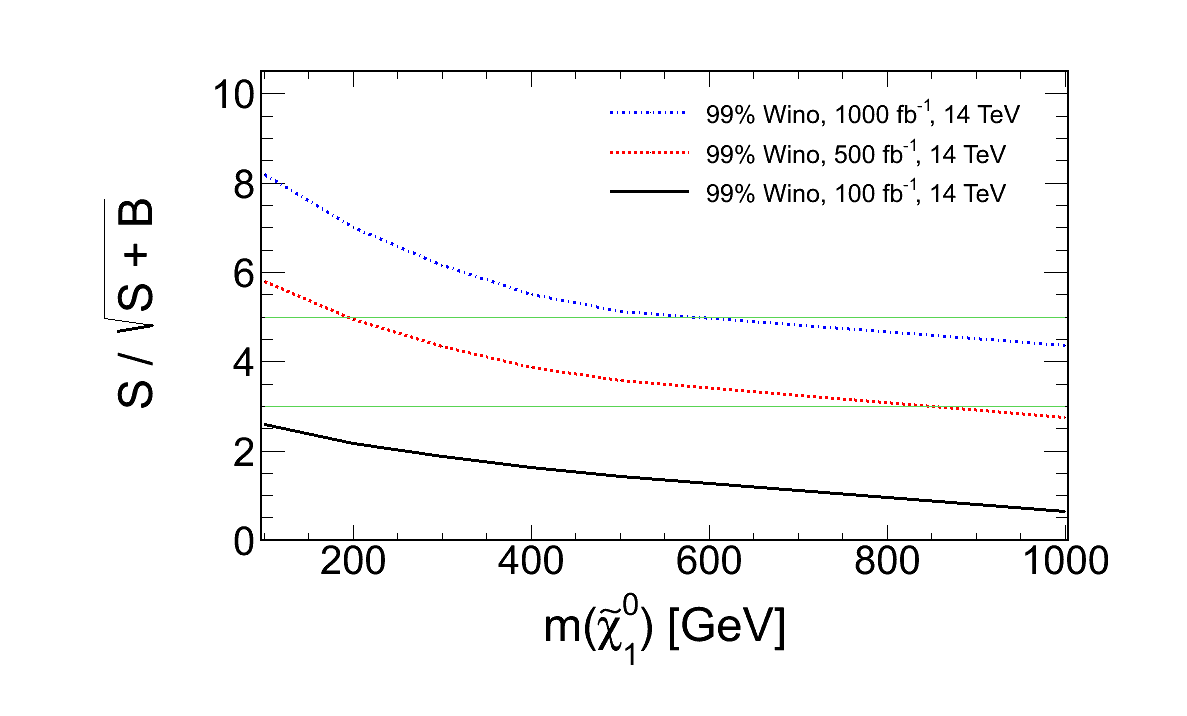}
\caption{Significance curves for the case where $\neuo{1}$ is $99\%$ Wino as a function of $m_{\neuo{1}}$ mass for different luminosities at LHC14. The green lines correspond to $3\sigma$ and $5\sigma$ significances~\cite{Delannoy:2013ata}.}
\label{SignificanceVsMassVsLumi_versionD}
\end{figure}

We plot the significance as a function of $\neuo{1}$ mass  in Fig. \ref{SignificanceVsMassVsLumi_versionD} using different luminosities at LHC14. 
The blue, red, and black curves correspond to luminosities of $1000, 500,$ and $100$ fb$^{-1}$, respectively. We observe that at $1000$ fb$^{-1}$, a significance of $5\sigma$ can be obtained up to a Wino mass of approximately $600$ GeV.  
We carry out the analysis by changing the jet energy scale and lepton energy scale by 20\% and 5\%, respectively.
We find the uncertainties in the significance to be 4\%.

In the case heavy sleptons, the subsequent decay of the second and third lightest neutralinos into the lightest neutralino occur via $Z$ bosonwhich  generates final states with $\geq$ $2$ jets, dileptons, and missing energy as in the light sleptons case. However, small branching fraction of ${\cal B}(Z \rightarrow ll)$ results in decreasing the discovery sensitivity, compared to the light slepton case. 

Determining the composition of $\neuo{1}$ for a given mass is very important in order to calculate the DM content. For example, if $\neuo{1}$ has a large Higgsino or Wino component, the annihilation cross section becomes large to fit the observed relic density for $m_{\neuo{1}}$ mass less than $\sim 1$ TeV for Higgsinos \cite{Allahverdi:2012wb} and $\sim 2.5$ TeV for Winos. On the other hand if $\neuo{1}$ is mostly Bino, the annihilation cross section is too small. In the first case one has under-abundance whereas in the second case one has over-abundance of DM. Both problems can be solved if the DM is non-thermal \cite{Allahverdi:2012gk} (in the case of thermal DM, addressing the over abundance  problem requires addition effects like resonance, coannihilation etc. in the cross section, while the under-abundance problem can be addressed by having multi-component DM \cite{Baer:2012cf}). If  $\neuo{1}$ is a suitable mixture of Bino and Higgsino, the observed DM relic density can be satisfied.   

Using Figs. \ref{xsectionColor} and \ref{Met_VBFDM_versionB}, we find that  varying  the rate and the shape of the $\met$ distribution it is possible to solve for the mass of $\neuo{1}$ as well as its composition in gaugino/Higgsino eigenstates. The VBF study described in this work was performed over a grid of input points on the $F - m_{\neuo{1}}$ plane (where $F$ is the Wino or Higgsino percentage in $\neuo{1}$). The $\met$ cut was optimized over the grid, and the $\met$ shape and observed rate of data were used to extract $F$ and $m_{\neuo{1}}$ which was then used to determine the DM relic density. 



\begin{figure}[!htp]
\centering
\includegraphics[width=3.5in]{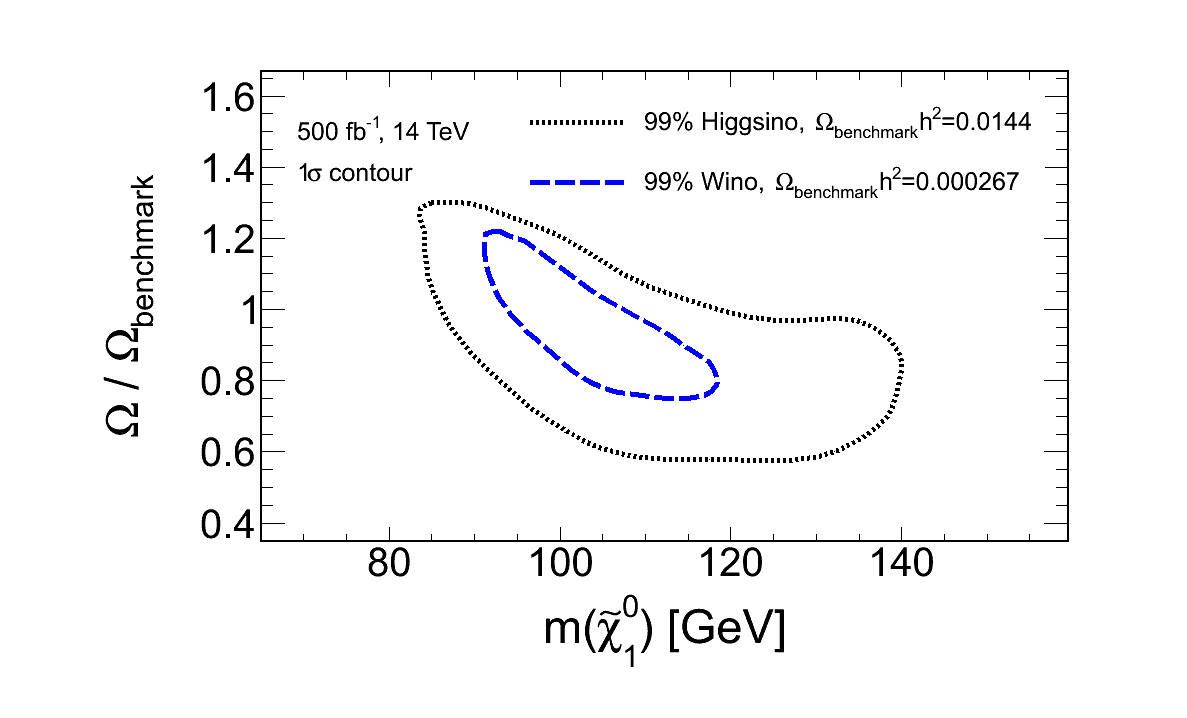}
\caption{Contour lines in the relic density-$m_{\neuo{1}}$ plane for $99\%$ Wino (blue dashed) and $99\%$ Higgsino (grey dotted) DMs expected with $500$ fb$^{-1}$ of luminosity at LHC14. The relic density is normalized to its value at $m_{\neuo{1}} = 100$ GeV~\cite{Delannoy:2013ata}.}
\label{OmegaVsMLSP_WinoAndHiggsino_versionD}
\end{figure}

In Fig. \ref{OmegaVsMLSP_WinoAndHiggsino_versionD}, we show the case of $99\%$ Higgsino and $99\%$ Wino using $1\sigma$ contour  on the relic density-$m_{\neuo{1}}$ plane for $500$ fb$^{-1}$ luminosity at LHC14. The relic density is normalized to the benchmark value $\Omega_{\rm benchmark}$, which is the relic density for $m_{\neuo{1}} = 100$ GeV.  For the Wino case, we find that the relic density can be determined within $\sim 20\%$, while for the Higgsino case it can be determined within $\sim 40\%$. We need larger luminosity for higher values of $m_{\neuo{1}}$ to achieve these results. We note we have not evaluated the impact of any degradation in $\met$ scale, linearity and resolution due to large pile-up events. It is important to revisit with the expected performance of upgraded ATLAS and CMS detectors. 

\section{Monojet}
The DM scattering off a nucleon is described by  $q \chi \to q \chi$  and the  collider productions of DM particles  can be written as $q\bar q \to \chi \bar{\chi}$.  A major theoretical and experimental effort is ongoing using effective interaction approach 
to describe the interactions of the DM particle with the
standard model (SM) particles.  
The signal becomes detectable after one attaches  either a gluon, photon, $Z$, or
a $W$ boson and the signal contains large missing energies associated with a pair of DM productions in association with jets, photons,
or leptons (from $W$ or $Z$ decays)~\cite{bib:othermonojet}. 

\begin{figure}[!htp]
\centering
\includegraphics[width=2.5in]{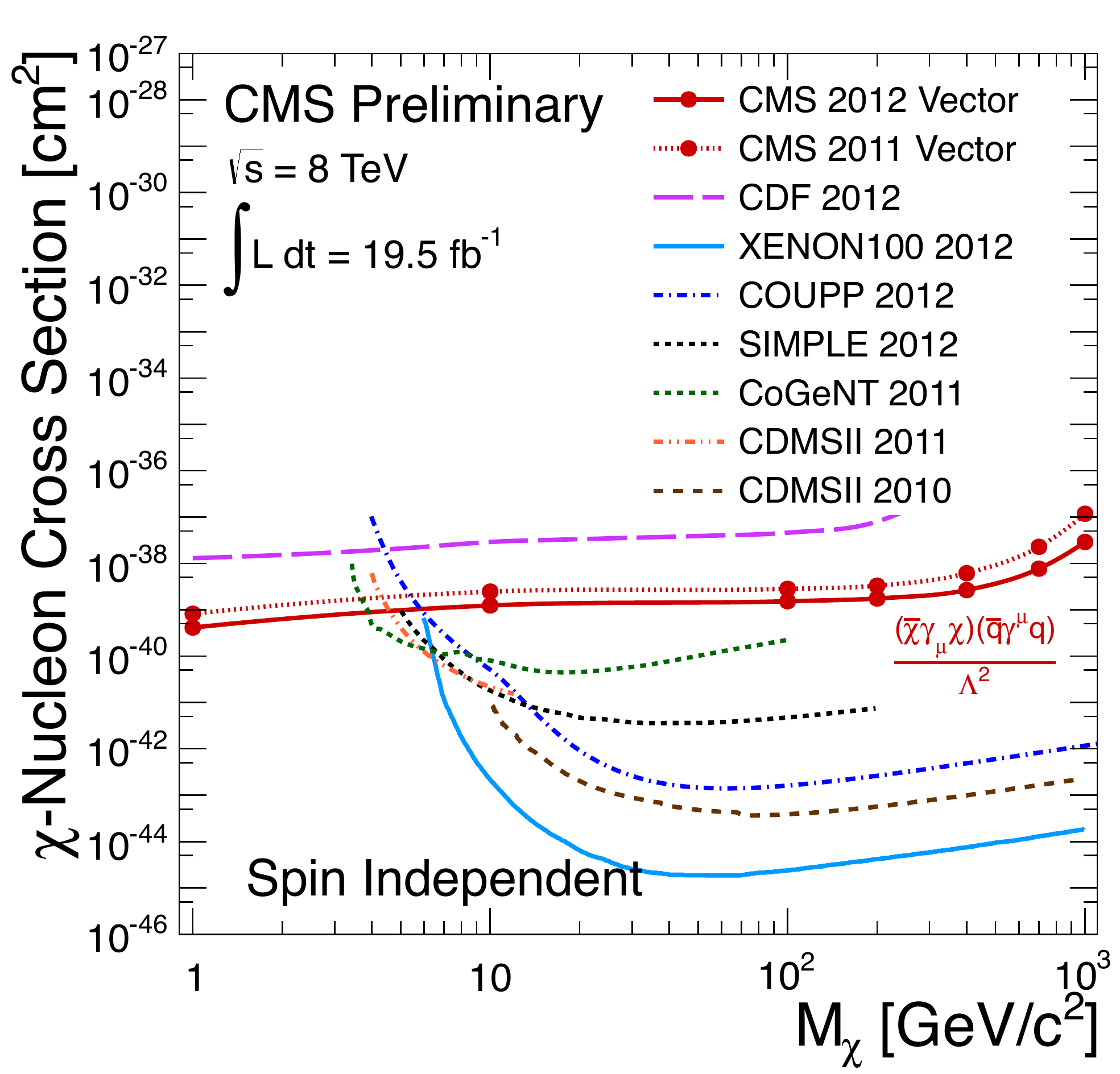}
\includegraphics[width=2.5in]{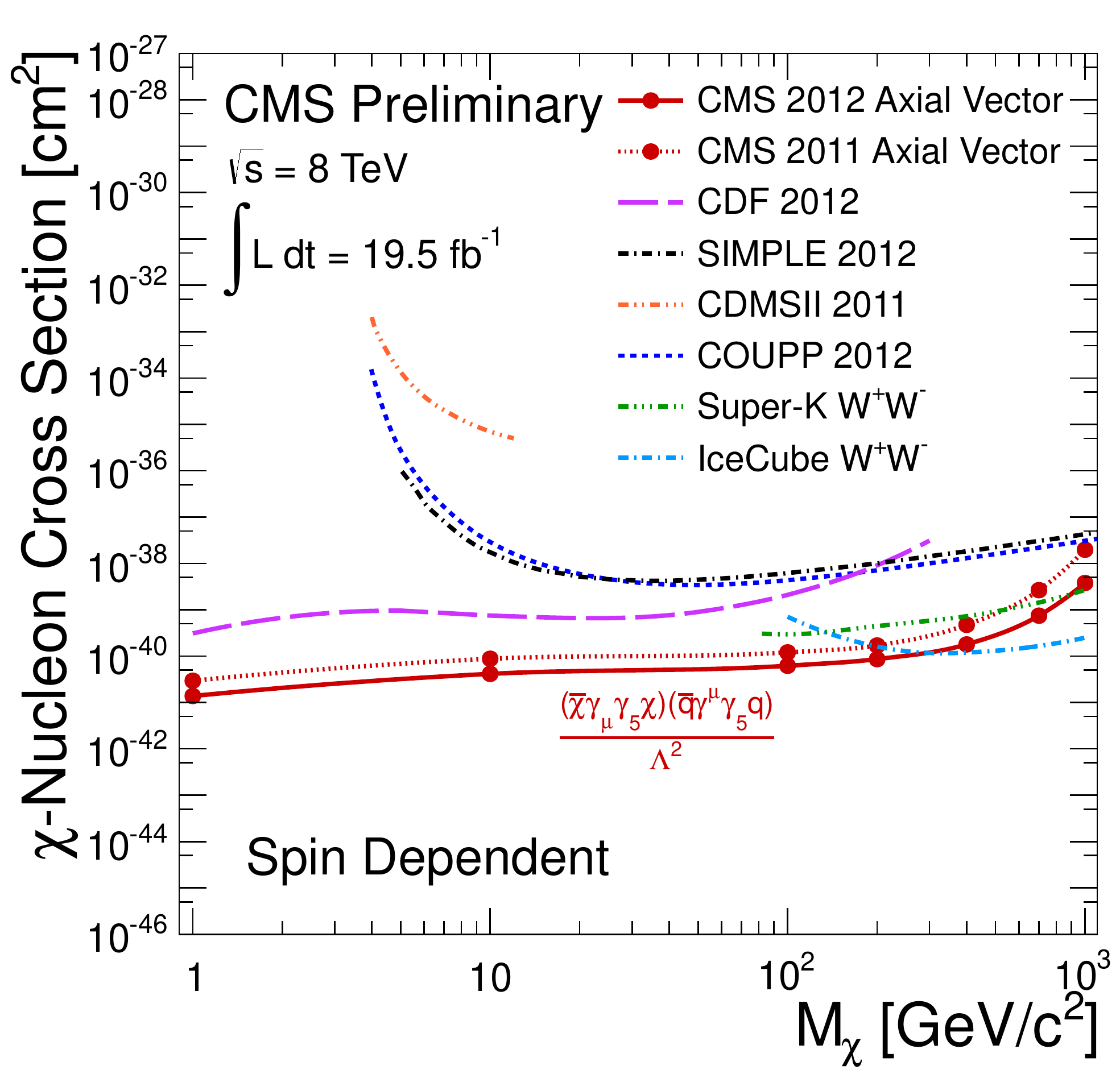}
\caption{Collider and various direct and indirect detection experiment constraints on spin indpenedent and spin dependent effective operators are shown~\cite{cms}.}
\label{monojetconstrain}
\end{figure}

For example, if we attach a gluon or a photon to a quark leg of the operators, $(\bar \chi
\chi)(\bar q q)$, we get a monojet or a monophoton plus missing energy
event. The monojet and monophoton production from the LHC
\cite{atlas,cms} and the direct detection of DM can be used to understand the effective operators.
In Fig.~[\ref{monojetconstrain}], we show collider and various direct and indirect detection experiment constraints on spin indepenedent effective operator ${{1\over{\Lambda^2}}} 
 \left( \bar \chi \gamma^\mu \chi \right)  
  \left( \bar q \gamma_\mu q \right)$  and spin dependent effective operator ${{1\over{\Lambda^2}}} 
  \left( \bar \chi \gamma^\mu \gamma^5 \chi \right) 
 \left( \bar q \gamma_\mu \gamma^5 q \right)$.

The monojet also arises naturally in the context of  minimal nonthermal DM model with a 1-GeV DM candidate, which naturally explain baryongensis~\cite{Allahverdi:2013mza}. Since the light DM is not parity-protected, it can be singly produced at the LHC.
This leads to large missing energy associated with an energetic jet whose transverse momentum distribution is featured by a Jacobian-like shape. 
The monojet, dijet, paired dijet and 2 jets + missing energy
channels are studied~\cite{Dutta:2014kia}. 

The interaction Lagrangian is given as,
\be {\cal L}_{int}=
\lambda_1^{\alpha,\rho\delta}\epsilon^{ijk}X^{}_{\alpha,i} \bar{d}^c_{\rho,j}{\textbf{P}}_R d_{\delta,k}+ \lambda_2^{\alpha,\rho} X^*_{\alpha} \bar{n}_{{DM}} {\textbf{P}}_R u_{\rho} +  \rm{C.C.}\label{eq:L_int}
\end{equation}
where $d^c$ is the charge-conjugate of the Dirac spinor. ${\textbf{P}}_R$ is the right-handed projection operator. $X$s are iso-single color triplet scalars with hypercharge 4/3 and $n_{DM}$ is a SM singlet which is DM candidate in this model. For the indices, $\rho,\delta=\{1,2,3\}$ denote the three quark generations, and $i,j=\{1,2,3\}$ are the SU(3) color indices. Successful baryogenesis requires more than one new scalar, thus $\alpha=1,2$ denotes for a minimal case with two $X$ fields.
\begin{figure}[h]
\centering
\includegraphics[scale=0.7]{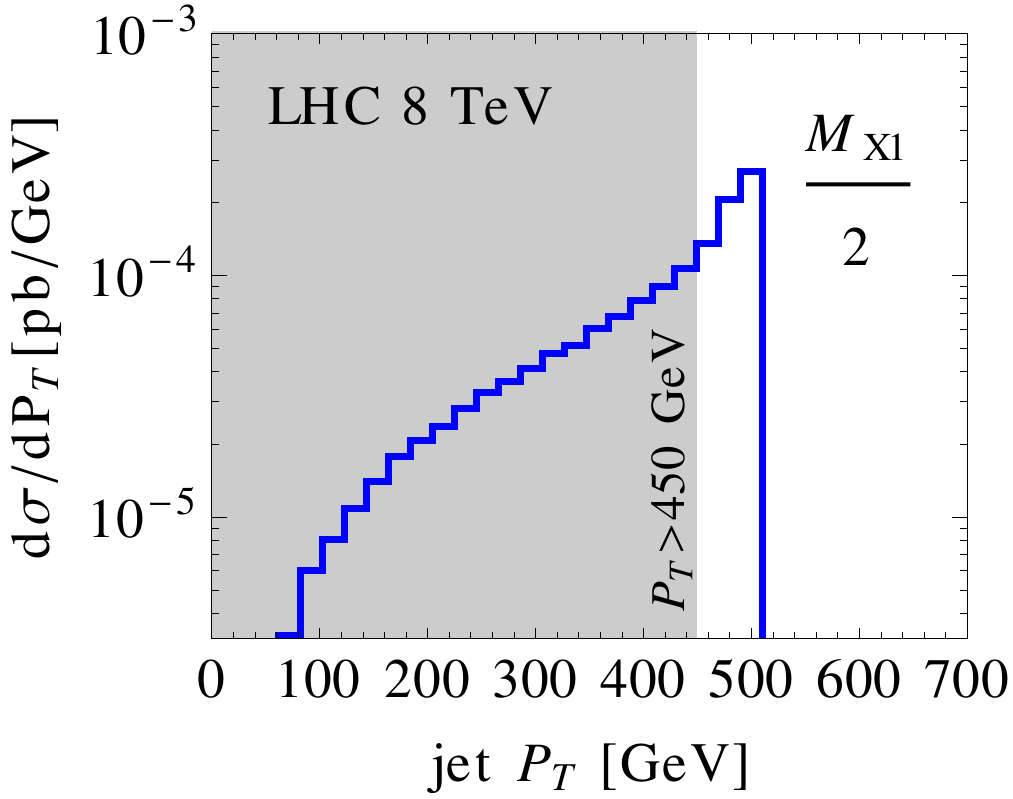}
\caption{Monojet $p_T$ distribution for $M_{X1}$=1 TeV~\cite{Dutta:2014kia}.}
\label{fig:pt_distr}
\end{figure}

In Fig.~\ref{fig:pt_distr} we show the distribution of jet transverse momentum is featured by two Jacobian-like  peaks near one half of the resonance energy $\sqrt{\hat{s}}=M_{X1}$ and $\sqrt{\hat{s}}=M_{X2}$. The transverse mass of the leading jet $p_{T}$ and $\met$ infers the mass of $X1$ and provide a maximal signal significance.

\section{Stop Squark and  the Lightest Neutralino}
Light top squark searches in most cases are  carried out where the  lightest neutralino ($\neuo{1}$) is mainly a Bino and the second lightest neutralino ($\neuo{2}$) mainly a Wino. 
In such a scenario, the lightest top squark ($\tilde{t}_1$) decays to $\neuo{1}$ and a top ($t$) quark at a branching fraction (${\cal B}$) of nearly $100\%$.
There have been many approaches in the literature to probe the stop squark in fully hadronic final state of
events  \cite{Dutta:2012kx, Plehn:2011tg}.
Reference  \cite{Graesser:2012qy} studied the  $\tilde{t}_1$  decay in the scenario  where $\neuo{1}$  and 
the lightest chargino ($\tilde\chi^\pm_{1}$) are purely Higgsino and a new variable, topness, was introduced  to identify the top squark from the top quark backgrounds.

If $\neuo{1}$ is primarily a Bino, its relic density is usually large since the annihilation cross-section is usually smaller than the required thermal annhilation rate
 $3\times 10^{-26}$ cm$^3$/sec. It is possible to obtain the correct relic density if the $\neuo{1}$ is a mixture of Bino and Higgsino \cite{ArkaniHamed:2006mb}, while having $\neuo{2}$ and $\neuo{3}$ as primarily Higgsinos. If $\neuo{1}$ is primarily a Wino, the annihilation into $W^{+}W^{-}$ final states is in tension with Fermi data for Wino mass below $\sim 250$ GeV \cite{fermiKoushiappas}.
In such a case,  we can have the following scenarios\cite{Dutta:2013sta}: 

\be
m_{\tilde t_1}> m_{\tilde\chi^0_3}, m_{\tilde\chi^0_2},m_{\tilde\chi^\pm_1} >m_{\tilde l}>m_{\tilde\chi^0_1} \,\,. 
\end{equation}
The possible  $\tilde{t}_1$ decay modes are:
\begin{eqnarray}
\tilde t_1& \rightarrow& t\ \tilde\chi^0_1 \\
\tilde t_1& \rightarrow& t\ \tilde\chi^0_2     \rightarrow t\  l^{\mp} \tilde l^{(\ast)\pm} \rightarrow t\ l^{\mp}  l^{\pm}  \tilde\chi^0_1,\\
\tilde t_1&\rightarrow&  b\  \tilde\chi^\pm_1\rightarrow b\  l\bar{\nu}\tilde\chi^0_1\,\, {(\rm or}\ b\ q\bar{q}^{\prime} \tilde\chi^0_1)  \\
\tilde t_1& \rightarrow& b\ \chpm{2} \rightarrow b\ Z\chpm{1} 
\end{eqnarray}

\noindent The last mode is allowed when the $\chpm{2}$ is lighter than $\tilde t_1$.

It is clear that one obtains an edge in the dilepton invariant mass distribution as well as $Z$-peak depending on the size of ${\cal B}(\tilde{t}_1 \rightarrow b \chpm{2})$ value.
A mass spectrum at our benchmark point is displayed in Table~\ref{benchparameters0}. 
\begin{table}[!htp] 
\caption{SUSY masses (in GeV) at the benchmark point~\cite{Dutta:2013sta}. }
\label{benchparameters0}
\begin{center}
\begin{tabular}{c c c} \hline \hline
 Particle  & Mass (GeV)      & ${\cal B}$  \\ \hline  \\[-.1in]
  $\tilde t_1$  & $500$     &  $17\%$ ($t\neuo{2}$), $22\%$ ($t\neuo{3}$), $8\%$ ($t\neu{1}$) \\
   &    &                            $53\%$ ($b\tilde\chi^\pm_1$)\\\hline \\[-.1in]
  $\neu{2} $ & $175$       &  $100\%$ ($l\tilde l$)  \\
  $\neu{3} $ & $176$       & $88\%$ ($l\tilde l$)   \\
$\chpm{1} $ &  $164$    & $22\%$ ($l \nu\neuo{1}$)\\
  $\tilde l $ & $144$       & $100\%$ ($l\neuo{1}$)  \\
  $\neuo{1}$  & $113$  & \\    \hline \hline
\end{tabular}
\end{center}
\end{table}

In the benchmark scenario, the mass difference between $\neuo{2,3}$ and $\neuo{1}$ is around $63$ GeV, and thus an edge in the dilepton invariant mass distribution is expected around this value.  
The final state of $2$ jets $+$ $2$ leptons $+$ $\met$ events  arises mostly from a combination of  the $\tilde{t}_1 \rightarrow b\tilde{\chi^{\pm}_1}$ and $\tilde{t}_1 \rightarrow t \neuo{2,3}$  decays. 
If both top squarks decay into a $b$ and a $\chi^{\pm}_1$, then  $2b$ + $2l$ + $\met$ events are expected. We apply the following cuts:
\begin{enumerate}
\item[($i$)]
At least $2$ isolated  leptons ($e$ or $\mu$) with $\pt  > $  20 and 10  GeV in $|\eta| \, < \, 2.5$, where the isolation is defined as  $\sum p_{T }^{\rm track} \, < \, 5$ GeV with $\Delta R = 0.4$;
\item[($ii$)]
At least $2$ jets with $\pt > 30$ GeV in $|\eta| \, < \, 2.5$;
\item[($ii$)]
At least 1 $b$-tagged jet with $\pt > 30$ GeV in $|\eta| \, < \, 2.5$;
\item[($iii$)] $\met \, > \, 150$ GeV;
\item[($iv$)]  $\HT \, > \, 100$ GeV.
\end{enumerate}

At this stage, the dominant SM background is $\ttbar$ events.
Opposite-sign same-flavour (OSSF) dileptons  arising from the $\neuo{2}$ decay  are kinematically correlated and its dilepton invariant mass distribution is expected to have an edge given by

\be
M^{\rm edge}_{ll} \,\, \sim \,\, m_{\neuo{2}} - m_{\neuo{1}} \,\,.
\end{equation}

The OSSF dilepton mass distribution from $\ttbar$ events are modelled by the dilepton distribution of opposite-sign different-flavour (OSDF) dilepton events. 
The OSSF dilepton mass distribution  from supersymmetric combinatoric background ($i.e.$, uncorrelated leptonic pairs) are also  modelled by OSDF dilepton mass distribution. 
This leads to adoping subtracting OSDF distributiuon from OSSF distribution.
The ` benchmark events would arise in an excess in OSSF$-$OSDF dilepton mass distribution.

\begin{figure}[!htp] 
\centering
\includegraphics[width=3.0in]{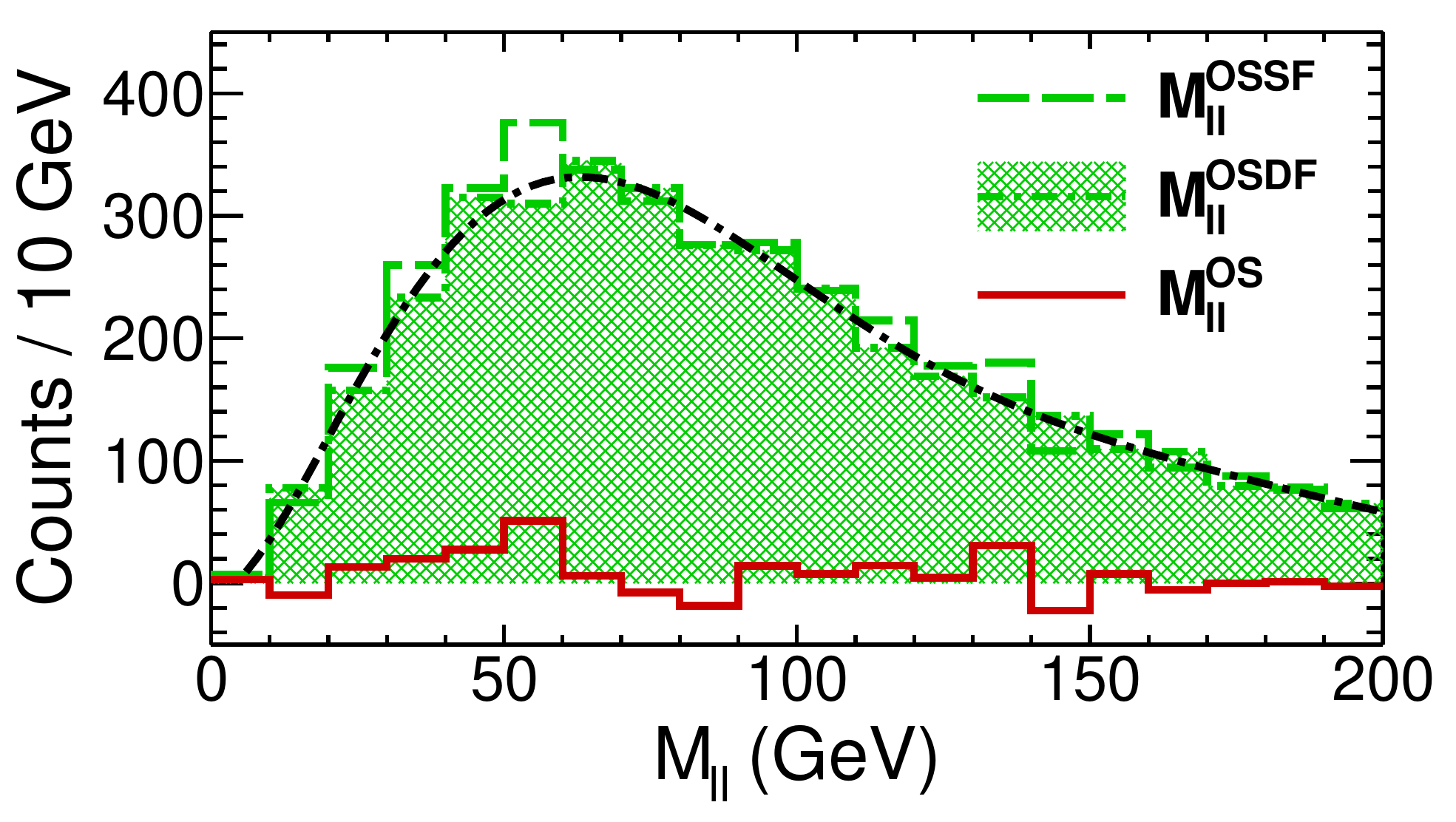}
\caption{The dilepton invariant mass distributions for $t \overline{t} + (0-4)$ jets background and the benchmark point in Table~\ref{benchparameters0} are displayed for  $30$ fb$^{-1}$ luminosity. The unshaded histogram shows the $M^{OSSF}_{ll}$ distribution, while the shaded histogram shows the $M^{OSDF}_{ll}$ distribution, which is fitted with the dot-dashed curve. The solid curve shows the subtracted distribution~\cite{Dutta:2013sta}.}
\label{MllOS_stopChange_susy1a1_lum30_sub_DFfitting_10p}
\end{figure}

\begin{figure}[!htp] 
\centering
\includegraphics[width=3.0in]{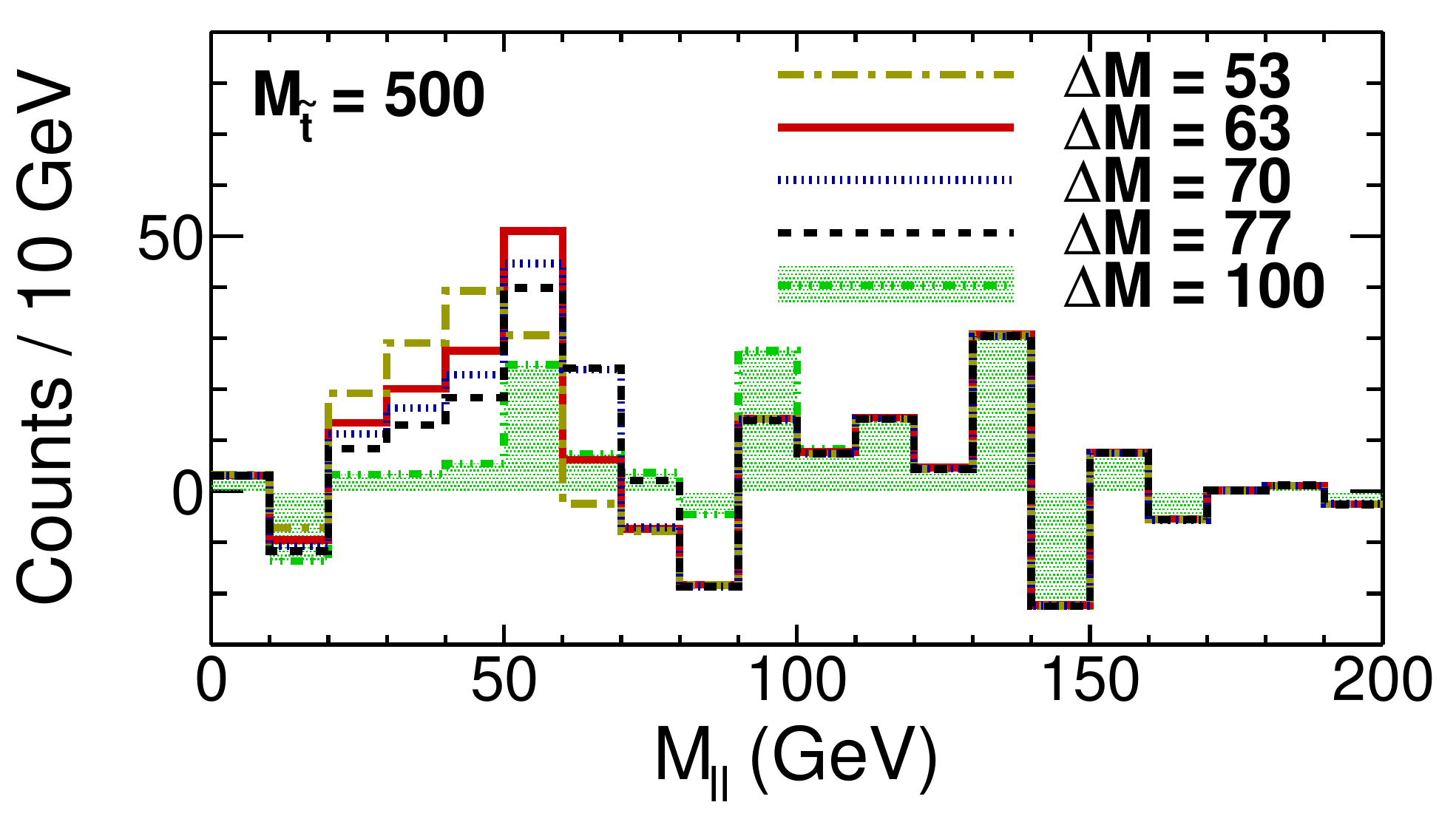}
\caption{The dilepton invariant mass distribution as $\neuo{2} - \neuo{1}$ is varied, for $\tilde{t}_1 = 500$ GeV and $\neuo{1} = 113$ GeV for  $30$ fb$^{-1}$ luminosity~\cite{Dutta:2013sta}.}
\label{MllOS_edgeShift_susyabcttjf_lum30_sub_DFfitting_10p}
\end{figure}

\begin{figure}[!htp] 
\centering
\includegraphics[width=3.0in]{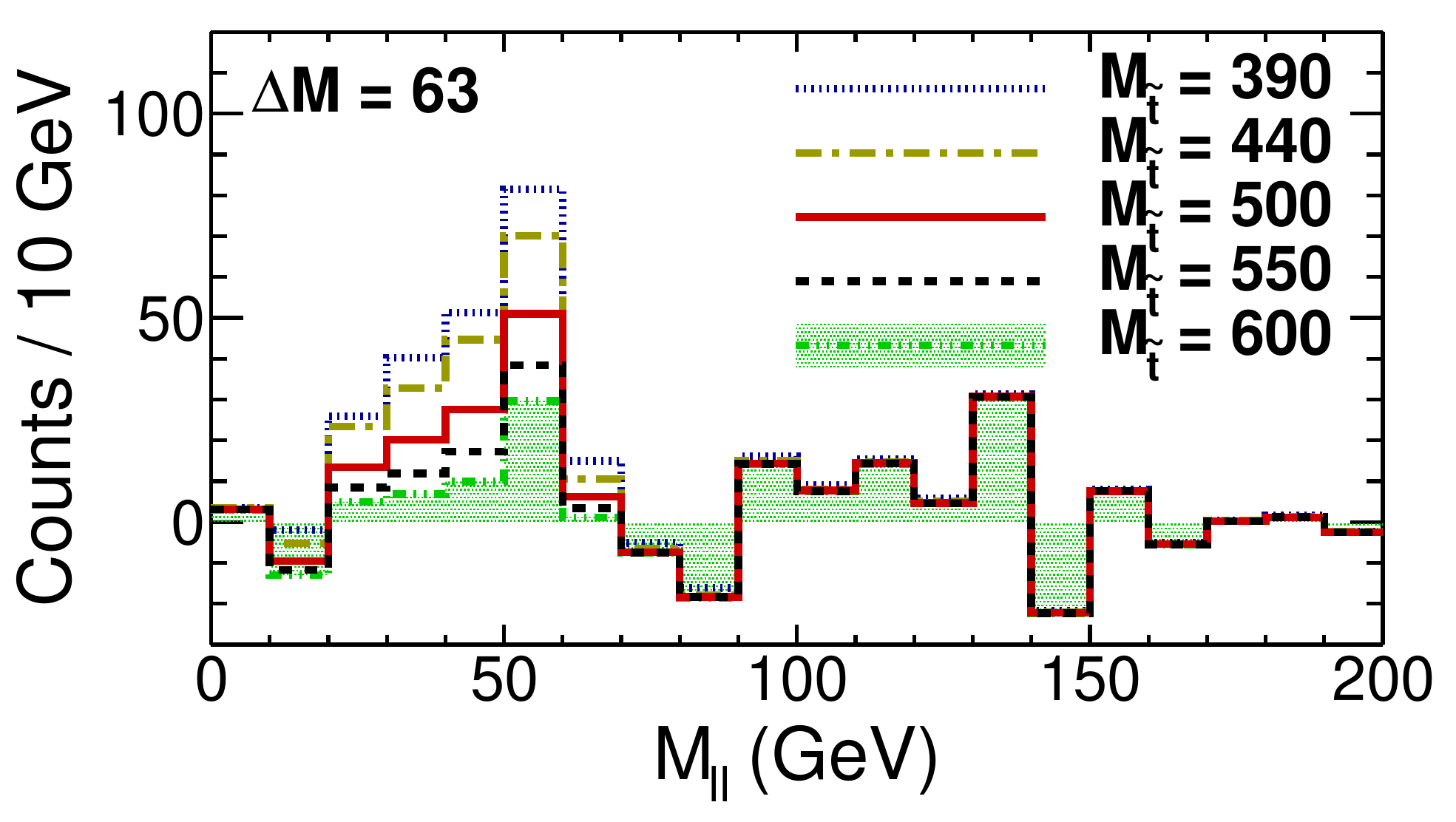}
\caption{The dilepton invariant mass distribution as $\tilde{t}_1$ mass is varied, all other masses remaining at the benchmark value in Table~\ref{benchparameters0} for  $30$ fb$^{-1}$ luminosity~\cite{Dutta:2013sta}.}
\label{MllOS_stopChange_susyabcttjf_lum30_sub_DFfitting_10p}
\end{figure}

 The OSDF dilepton  mass distributions for the SUSY benchmark point in Table \ref{benchparameters0} along with SM $\ttbar$ + (0-4)~jets background is shown in shaded histogram in Fig.~\ref{MllOS_stopChange_susy1a1_lum30_sub_DFfitting_10p}, while its OSSF distribution (blank histogram)  is overlayed. A clear edge is seen at around 63 GeV for  30 fb$^{-1}$ luminosity.
Fig.~\ref{MllOS_edgeShift_susyabcttjf_lum30_sub_DFfitting_10p} shows the flavor subtracted distributions at $\Delta M$ = $m_{\neuo{2}} -  m_{\neuo{1}}$ =  $53, 63, 70, 77$ and $100$ GeV for $\tilde{t}_1 = 500$ GeV,  $\neuo{1} = 113$ GeV and
$m_{\neuo{3}} \sim m_{\neuo{2}}$. The dilepton mass distribution edge for all these mass differences can be seen clearly.
Fig.~\ref{MllOS_stopChange_susyabcttjf_lum30_sub_DFfitting_10p} shows the flavor subtracted distribution for $\ttbar$ + (0-4)~jets background plus signal events  for $m_{\tilde{t}}$  = $390, 440, 500, 550$ and $600$ GeV, with $m_{\neuo{1}} = 113$ GeV and $m_{\neuo{3}} \sim m_{\neuo{2}}$ = $175$ GeV. The edge of the dilepton distribution for $m_{\tilde{t}}$ mass  upto $550$ GeV can  be distinguished from the background for  $30$ fb$^{-1}$ luminosity.

An excess for each top squark mass case in  Fig. \ref{MllOS_stopChange_susyabcttjf_lum30_sub_DFfitting_10p} is evaluated  in terms of significances (${\cal S}$) in Table \ref{stopsignificance}. We show significances in  cases where  at least one of the jets required to be a b -jet and also in  cases without any b-jet requirement.
Here ${\cal S} = N_{\rm S} / \sqrt{N_{\rm S} + N_{\rm B}}$, where $N_{\rm S}$ and $N_{\rm B}$ are the number of OSSF dilepton events in range of $20$ GeV  $<  M_{ll}  <$  $70$ GeV  for signal (S) and background (B), respectively.

$N_{\rm B}$ is determined by fitting  the entire (susy plus $\ttbar$) OSDF dilepton distribution to a polynomial function and calculating the number of events in 20 GeV  $<  M_{ll}  <$  70 GeV. $N_{\rm S}$ is the the number of events in excess above $N_{\rm B}$. We find that the significance of the benchmark scenario for $m_{\tilde t_1} = 500$ GeV is above 3 $\sigma$ for $30$ fb$^{-1}$
\begin{table}[!htp] 
\caption{Significances (${\cal S}$) at $30$ fb$^{-1}$ for various $\tilde{t}_1$  masses  with $\neu{1} = 113$ GeV and $\neu{2},\neu{3} = 175$ GeV~\cite{Dutta:2013sta}.}
\label{stopsignificance}
\begin{center}
\begin{tabular}{c  c c} \hline \hline \\[-.1in]
   $\tilde{t}_1$  Mass      & ${\cal S}$ & ${\cal S}$   \\ 
                        (GeV)      & ($\geq 1$ $b$ )  &    \\ \hline  \\[-.1in]
$390$   &  $5.3$ & $6.3$  \\
$440$   & $4.6$  & $5.4$   \\
$500$ (benchmark)   &  $3.1$ & $3.5$  \\
$550$   &  $2.1$ & $2.3$  \\
$600$   &  $1.4$ & $1.4$  \\ \hline \hline

\end{tabular}
\end{center}
\end{table}

\section{Conclusion}
In conclusion, we have mentioned  four different ways of investigating DM at the LHC, e.g.,  (i) cascade decays of colored particles, (ii) VBF productions of non-colored particles, (iii) monojet and (iv) direct stop productions. 
The technique developed for cases (i) and (ii) are discussed in the context of SUSY models, but they can be applied to any model where we heavy color states and new particles with electroweak charges. 

In case (i) we discussed the production of the DM candidate, $\tilde\chi^0_1$, from the  cascade decays arising from   $\tilde q\tilde g$, $\tilde q \tilde q$, $\tilde g\tilde g$ etc. The end-points of various distributions~\cite{hinchliffe}, $M_{ll}$, $M_{\tau\tau}$, $M_{j l}$, $M_{j\tau}$, $M_{j\tau\tau}$, $M_{jW}$, $p_T$ of $e$, $\mu$ and $\tau$ etc. are needed to be measured in order to measure the masses of various particles (including $\tilde\chi^0_1$).  However, determining these observables requires removal of combinatoric backgrounds from the SUSY and SM processes. The importance of bi-event subtraction techniques to remove these backgrounds is shown and the DM content is determined in the simple models, e.g., nuSUGRA, mSUGRA, minimal mirage mediation models etc.

In case (ii),  the charginos, neutralinos and sleptons which  enter into the DM content calculation are directly produced by VBF. The VBF production is characterized by the presence of two energetic jets in the forward direction in opposite hemispheres. The central region is relatively free of hadronic activity, and provides a potential probe of the supersymmetric EW sector, with the SM and $t \overline{t}+$ jets background under control.  The cases of pure Wino, pure Higgsino, and mixed Bino-Higgsino DM have been studied using VBF productions in the $2j + \met$ final state at 14 TeV.  By optimizing the $\met$ cut for a given $m_{\neuo{1}}$, one can simultaneously fit the $\met$ shape and observed rate in data to extract the mass and composition of $\neuo{1}$, and hence solve for the DM relic density. At an integrated luminosity of $1000$ fb$^{-1}$, a significance of $5\sigma$ can be obtained up to a Wino mass of approximately $600$ GeV. The relic density can be determined to within $20\% \, (40\%)$ for the case of a pure Wino (Higgsino) for $500$ fb$^{-1}$ at LHC14, for $m_{\neuo{1}} = 100$ GeV. 

In case (iii), monojet final state made of single jet and missing transverse energy is discussed. The DM particles can be pair produced along  with a jet from initial state radiation to give rise to this signal which can be used to study interaction between the DM particle and SM particles.  The collider and  the DM-nucleon scattering cross sections  for both spin-independent and spin-dependent interaction models can be utilized simultaneousely to investigate the scale $\Lambda$ associated with the interaction.  

The monojet signal also can arise when DM is singly produced along with a jet.  The collider phenomenolgy of such a model which explains baryogenesis along with a 1 GeV DM is discussed. 
In this type of models, the large missing energy is associated with an energetic jet whose transverse momentum distribution is featured by a Jacobian-like shape.

In case (iv), light top squark (stop) pair production at the LHC is used to probe the nature of $\tilde\chi^0_1$. When the lightest neutralino is a mixture of Bino and Higgsino in order to satisfy the thermal relic density, the lightest stop decays mostly into $(i)$ a top quark plus the second or third lightest neutralino, and $(ii)$ a bottom quark plus the lightest chargino, instead of the stop decay scenario that is mostly searched, where the lightest stop decays $100\%$ into a top plus the lightest neutralino. The subsequent decay of the second and third lightest neutralinos into the lightest neutralino via an intermediate slepton  or $Z$ boson generates final states with $\geq$ $2$ jets, dileptons, and missing energy. We find that the dilepton mass distribution after subtracting the opposite sign different flavor (OSDF) distribution from the opposite sign same flavor (OSSF) distribution shows a clear edge in the case of  light sleptons.

\Acknowledgements
I would like to thank my collaborators Richard Arnowitt, Adam Arusano, Andrea Delannoy, Will Flanagan, Yu Gao, Tathagata Ghosh, 
Alfredo Gurrola, Will Johns, Teruki Kamon, Nikolay Kolev, Abram Krislock, Eduardo Luiggi, Andrew Melo, Paul Sheldon, Paul Simeon, Kuver Sinha, Dave Toback, Kechen Wang and Sean Wu for the works related to this review. This work is supported in parts by DOE grant DE-FG02-13ER42020.

\end{document}